\documentclass[fleqn,usenatbib]{mnras}

\usepackage{newtxtext,newtxmath}

\usepackage[T1]{fontenc}
\usepackage{ae,aecompl,times}

\usepackage{graphbox,graphicx}	
\usepackage{amsmath}	
\usepackage{multirow}
\usepackage[normalem]{ulem}
\usepackage{xcolor}
\usepackage{bm}
\usepackage{lipsum}

\title[Applying a calibrated star formation subgrid model in cosmological simulations]{Applying a star formation model calibrated on high-resolution interstellar medium simulations to cosmological simulations of galaxy formation}

\author[J.~D.~Burger et al.]{%
\parbox{0.99\textwidth}
{%
Jan D.~Burger$^{1}$\thanks{E-mail: burger@mpa-garching.mpg.de},
Volker Springel$^{1}$, 
Eve C.~Ostriker$^{2, 3}$,
Chang-Goo Kim$^{2}$,
Sarah M.~R.~Jeffreson$^{4}$,\\
Matthew C.~Smith$^{1}$,
R\"{u}diger Pakmor$^{1}$,
Sultan Hassan$^{5}$,
Drummond Fielding$^{6}$,
Lars Hernquist$^{4}$,\\
Greg L.~Bryan$^{7, 8}$,
Rachel S.~Somerville$^{8}$,
Jake S.~Bennett$^{4}$, and
Rainer Weinberger$^{9}$
}
\\
\\%
$^{1}$Max-Planck-Institut f\"ur Astrophysik, Karl-Schwarzschild-Str. 1, D-85748, Garching, Germany\\%
$^{2}$Department of Astrophysical Sciences, Princeton University, 4 Ivy Lane, Princeton, NJ 08544, USA\\%
$^{3}$Institute for Advanced Study, 1 Einstein Drive, Princeton, NJ 08540, USA\\%
$^{4}$Center for Astrophysics | Harvard \& Smithsonian, 60 Garden Street, Cambridge, MA 02138, USA\\%
$^{5}$Center for Cosmology and Particle Physics, Department of Physics, New York University, 726 Broadway, New York, NY 10003, USA\\%
$^{6}$Department of Astronomy, Cornell University, Ithaca, NY 14853, USA\\%
$^{7}$Department of Astronomy, Columbia University, 550 West 120th Street, New York, NY 10027, USA\\%
$^{8}$Center for Computational Astrophysics, Flatiron Institute, 162 5th Ave, New York, NY 10010, USA\\%
$^{9}$Leibniz Institute for Astrophysics Potsdam (AIP), An der Sternwarte 16, 14482 Potsdam, Germany
}

\date{Accepted XXX. Received YYY; in original form ZZZ}

\pubyear{2024}

\begin{document}
\label{firstpage}
\pagerange{\pageref{firstpage}--\pageref{lastpage}}
\maketitle

\begin{abstract}
  Modern high-resolution simulations of the interstellar medium (ISM) have shown that key factors in governing star formation are the competing influences of radiative dissipation, pressure support driven by stellar feedback, and the relentless pull of gravity. Cosmological simulations of galaxy formation, such as IllustrisTNG or ASTRID, are however not able to resolve this physics in detail and therefore need to rely on approximate treatments. These have often taken the form of empirical subgrid models of the ISM expressed in terms of an effective equation of state (EOS) that relates the mean ISM pressure to the mean gas density. Here we seek to improve these heuristic models by directly fitting their key ingredients to results of the high-resolution TIGRESS simulations, which have shown that the dynamical equilibrium of the ISM can be understood in terms of a pressure-regulated, feedback modulated (PRFM) model for star formation. Here we explore a simple subgrid model that draws on the PRFM concept but uses only local quantities. It accurately reproduces PRFM for pure gas disks, while it predicts slightly less star formation than PRFM in the presence of an additional thin stellar disk. We compare the properties of this model with the older Springel \& Hernquist and TNG prescriptions, and apply all three to isolated simulations of disk galaxies as well as to a set of high-resolution zoom-in simulations carried out with a novel ``multi-zoom'' technique that we introduce in this study. The softer EOS implied by TIGRESS produces substantially thinner disk galaxies, which has important ramifications for disk stability and galaxy morphology. The total stellar mass of galaxies is however hardly modified at low redshift, reflecting the dominating influence of large-scale gaseous inflows and outflows to galaxies, which are not sensitive to the EOS itself. 
\end{abstract}




\section{Introduction}
\label{sec:intro}

Cosmological hydrodynamic simulations have become a powerful theoretical tool to investigate the formation of galaxies in the currently favoured $\Lambda$CDM cosmology, as well as in possible alternatives. They are nowadays typically carried out either in homogeneously sampled periodic boxes, or by focusing the computational effort into a single object embedded in a more coarsely resolved background using the so-called zoom-in technique \citep[see][for a review]{Vogelsberger2020}. Examples of influential recent calculations of the box-type include Illustris \citep{Vogelsberger2014}, EAGLE \citep{Schaye2015}, Magneticum \citep{Dolag2016}, HorizonAGN \citep{Dubois2016}, IllustrisTNG \citep{Springel2018}, SIMBA \citep{Dave2019}, or ASTRID \citep{Ni2022}, and of the zoom-in type, FIRE \citep{Hopkins2014}, Auriga \citep{Grand2017}, LYRA \citep{Gutcke2021}, or Vintergatan \citep{Agertz2021}, among others.

While all of these projects, particularly the zoom-in ones, push for ever better resolution, they are still far away from resolving the star formation processes in the multi-phase interstellar medium in an ab-initio fashion, let alone their regulation through energetic feedback from the radiation fields produced by stars,  stellar winds, and the violent thermonuclear explosions of stars known as supernovae. Cosmological simulations that cover representative volumes of the universe cannot yet form galaxies one star at a time, and cannot follow individual supernova explosions. While some simulations of individual low mass  galaxies start to make at least the latter possible \citep{Hu2016, Emerick2019, Gutcke2021}, it will be some time before this can be achieved in more massive galaxies, and it is questionable whether this can be reached in large cosmological volumes within the next decades. Simulation studies of full galaxy populations, as needed in particular for cosmological applications, will therefore have to continue to live  for the foreseeable future with comparatively coarse resolution. This in turn makes some kind of sub-grid treatment unavoidable.

The term `sub-grid' refers to the general concept that an approach is needed to account for physics that is spatially unresolved in a cosmological simulation but that nevertheless affects resolved scales. How such a `closure' is implemented at a technical level varies in different simulation methodologies, and there is no universally accepted strategy for it.

\citet[][hereafter SH]{Springel2003} have argued that a sub-grid model for the ISM is best formulated in terms of an analytic model for the substructure expected in a computational resolution element that represents the ISM. They augmented the fluid equations with corresponding source functions and showed that the resulting terms could be summarized by assigning each cell a mean effective pressure and mean star formation rate expected from the model. The advantage of this approach is that the results should in principle be insensitive to the numerical resolution employed and the exact hand-over scale between resolved flow and subgrid model, thereby making it possible to obtain numerically converged results. If this is achieved, the uncertainties in the results become dominated by the adopted physics model, while the uncertainties due to numerical resolution can be reduced to any desired level. Also, the parameters describing the subgrid model do not need to be changed when varying the resolution  \citep{Marinacci2014}. This type of approach has been applied, for example, in the IllustrisTNG and Auriga projects, among others.

Another line of thinking is to resort to a more heuristic description of the feedback processes, implemented in a way that appears physically sensible at the available resolution. Here one views the physical and numerical resolution limitations as tightly coupled, thereby accepting that the numerical parameterization of the feedback modelling needs to be adjusted whenever the numerical resolution itself is changed. This calibration philosophy of the subgrid feedback physics has been applied, for example, in projects such as EAGLE \citep{Schaye2015},  APOSTLE \citep{Sawala2016}, or FLAMINGO \citep{Kugel2023}.

A further approach lies in avoiding the introduction of an explicit subgrid treatment altogether, in favour of accounting for as much physics as possible, initially regardless of whether or not it can be adequately resolved at the available numerical resolution. Some adjustments in the numerical implementation of the galaxy formation physics may however be incorporated to enhance model behaviour, but in general the philosophy is to try to reach sufficient numerical resolution such that the results stabilize at a (hopefully) physically correct prediction. The degree to which such `meso-scale' models succeed in practice arguably varies, and this also depends on the type of physical quantity that is examined. Simulation projects that follow some version of this philosophy include FIRE \citep{Hopkins2014}, SMUGGLE \citep{Marinacci2019} or EDGE \citep{Agertz2020}. Their minimum resolution requirements tend to be substantially higher than what can be afforded in large volume simulations.

For very large hydrodynamical cosmological simulations \citep[such as MillenniumTNG,][]{Pakmor2023} that are needed, in particular, to work out accurate predictions for the baryonic impact on various cosmological probes we therefore consider the explicit subgrid approach to be the most practical. But it is evident that the accuracy and reliability of this approach directly depend on the quality of the employed subgrid model. It therefore appears worthwhile to replace the coarse treatment employed thus far in simulations like IllustrisTNG with models that are physically better justified. Our approach for realizing this is to ultilize results from high-resolution simulations of star-formation, such as obtained in projects like SILCC \citep{Walch2015} or TIGRESS \citep{Kim2017}. These studies have made significant advances in recent years towards faithfully modelling a turbulent multi-phase ISM in which star formation and its regulation by supernova explosions and radiative feedback processes are computed essentially from first principles. Our strategy therefore is to formulate a subgrid model that makes direct use of these results, with the goal to eventually use it for next generation large-volume cosmological simulations of galaxy formation. This work is also meant to support an important goal of the `Learning the Universe' collaboration\footnote{The Learning the Universe collaboration is generously funded by the Simons Foundation and consists of a network of cosmologists, computational astrophysicists, and experts in machine learning techniques at various research institutions in the US and Europe (see also at \url{https://learning-the-universe.org}). Among other topics, the collaboration works on improved sub-grid models for the influence of stars and black holes on galaxy formation, on advanced methodologies to speed up forward modelling, and on novel simulation-based inference techniques.}, namely to reduce the systematic uncertainty in the predictions of simulations of cosmic structure formation so that they can be used more powerfully as a basis for reliable cosmological inference. 

In this work we make use of the results of the TIGRESS simulation suite analyzed in \citet{Ostriker2022}. According to their results, the ISM in disk galaxies follows a well-defined scaling relation between the central mid-plane pressure and the surface density of star formation. Moreover, the authors find a  power-law equation of state that relates time- and spatially-averaged properties of the star-forming ISM (such as total gas density and average local gas pressure), albeit with large scatter. In this paper, we shall first compare the TIGRESS implied relations to the SH and IllustrisTNG models, both for analytic plane-parallel gas layers and isolated galaxy models, thereby identifying some areas where we expect differences between the models in the outcome of galaxy formation simulations. We then follow this up in a set of  different `multi zoom-in' cosmological simulations that target different galaxy masses and are run to redshift $z=0$. To make the latter simulations easily possible, we also introduce a new methodology for constructing the corresponding multi zoom-in initial conditions, which in essence are simply multiple zoom-ins carried out concurrently within one simulation box. This approach allows one to make statements for the population statistics of galaxies much more conveniently than with the classic, one at a time zoom-in technique.

This paper is structured as follows. In Section~\ref{sec:methodology} we discuss different equation of state models and how the vertical equilibrium in a star-forming gaseous layer can be used to determine key scaling relations of the models. In Section~\ref{Secdisks}, we then turn to simulations of isolated galaxies, which we primarily use for verification of the numerical implementation of the models introduced earlier. We then turn to introducing our multi-zoom technique in Section~\ref{SecZooms}, where we also analyze results of several sets of simulations at different galaxy mass scales, comparing the outcomes of the SH, TNG, and TIGRESS models. Finally, we discuss our findings and conclude in Section~\ref{SecConc}. This work is part of the ``Learning the Universe'' collaboration, aiming to build next-generation cosmological simulations that \textbf{}incorporate improved prescriptions for star formation modeling.

\section{Methodology}
\label{sec:methodology}

\citet{Ostriker2022} have outlined a pressure-regulated, feedback-modulated (PRFM) theory of star-formation in the interstellar medium (ISM). A central concept in their analysis is the realization that the ISM's physical state responds to the confining gravitational field such that a quasi-steady state is established in which the SFR adjusts as needed to provide the required feedback to balance the gravitational forces. In the following, we first summarize the principal aspects of the vertical equilibrium of gaseous, star-forming sheets, and then recap the SH and TNG models for definiteness. We then introduce our new TIGRESS-inspired model and compare it to these older EOS realizations as well as to full PRFM model.

\subsection{Plane parallel subgrid models}

The gravitational potential $\Phi$ for a total mass density field $\rho_{\rm tot}$ fulfils Poisson's equation
 \begin{equation}
   \nabla^2 \Phi = 4 \pi G \rho_{\rm tot} ,
 \end{equation}
and for hydrostatic equilibrium, the pressure gradient is balanced by the gravitational force, i.e.
\begin{equation}
-\frac{\nabla P}{\rho_{\rm gas}} - \nabla\Phi = 0.   \label{eqnequil}
\end{equation}
We now assume a (thin) plane-parallel geometry stratified in the $z$-direction, meant to represent a piece of the gaseous disk of a galaxy, and for the moment neglect stellar and dark matter contributions to the density (i.e.~$\rho_{\rm tot} = \rho_{\rm gas}=\rho$). Also, we assume that the pressure depends only on gas density, $P=P(\rho)$, which we refer to as an effective equation of state.

We can then readily transform the above equations into a system of two coupled, first-order, ordinary differential equtions for $\rho$ and an auxiliary variable $q\equiv {\rm d}\Phi / {\rm d}z$:
\begin{align}
  \frac{{\rm d} q}{{\rm d}z}  & =  4 \pi G \rho,  \label{eqnvertical1}
  \\
  \frac{{\rm d} \rho}{{\rm d}z} & =  - q\, \frac{\rho}{{\rm d}P/{\rm
                                  d}\rho}.  \label{eqnvertical2}
\end{align}
Adopting boundary conditions $\rho(0) = \rho_0$ and $q(0)=0$ at the midplane, this can be straightforwardly integrated from the midplane out to some height $z$ above the plane of the disk. Every starting value $\rho_0$ for the central density defines a unique density profile $\rho(z)$ for the self-gravitating gaseous sheet, as well as its surface mass density
\begin{equation}
 \Sigma_{\rm gas} = \int_{-\infty}^\infty \rho(z) {\rm d}z.
\end{equation}
Note that with the specification of a density-dependent star-formation law, $\rho_{\rm SFR} = \rho_{\rm SFR}(\rho)$, the density profile likewise determines the surface density of star formation, $\Sigma_{\rm SFR}$.
 
\subsection{The Springel \& Hernquist model}

A simple explicit subgrid model for the interstellar medium (ISM), meant to facilitate numerically well posed results at comparatively low simulation resolution, was introduced by \citet{Springel2003} based on the conjecture that the ISM can be modelled as being composed of dense, cold clouds embedded in a tenuous hot medium, with both phases being roughly at pressure equilibrium. The clouds are assumed to be growing by radiative cooling, while they can be consumed by star formation or destroyed by supernova feedback and thermal conduction. It needs to be stressed that this old picture has little to do with modern ISM theory, where it has been recognized that the ISM is not composed of hydrostatic clouds but rather forms a supersonically turbulent medium that is characterized by the constant formation and dispersal of filament- and cloud-like mass concentration. But the SH model can still serve as a framework to motivate the salient point of the regulation of star formation through a balance between dissipative radiative losses and stellar feedback heating.

Under a number of simplifying assumptions, SH showed that the considered medium quickly evolves towards an equilibrium state where a mass fraction $x=\rho_c/\rho$ is in the cold clouds, and that this state can be characterized by an effective mean pressure given by
\begin{equation}
  P_{\rm eff}=(\gamma - 1) \,\rho\, [(1 - x) u_h + x\, u_c].   \label{eqnSH}
\end{equation}
Here $\gamma = 5/3$ is the adiabatic index, $\rho$ is the total gas density, $\rho_c$ is the mean density of cold clouds, $u_h$ denotes the thermal energy per unit mass of the hot phase, while $u_c$ is the thermal energy per unit mass of the cold clouds which is set to a constant value in the model. The thermal energy of the hot phase is given in the equilibrium state by
\begin{equation}
u_h = \frac{u_{\rm SN}}{A + 1} + u_c  , \label{eqnhot}
\end{equation}
where $u_{\rm SN}$ is a constant characterising the supernova feedback energy per unit mass of stars formed. The quantity $A$ describes the efficiency of cloud evaporation by supernovae, and has a density dependence of $A\propto \rho^{-4/5}$ \citep{McKee1977}. It is parameterized as
\begin{equation}
  A(\rho) = A_0\left(\frac{\rho}{\rho_{\rm th}}\right)^{-4/5}
\end{equation}
in the following, where $A_0$ is a constant, and $\rho_{\rm th}$ is the star formation threshold density. Only for densities $\rho > \rho_{\rm th}$, star formation is assumed to take place and the subgrid model is applied.  SH furthermore assumed that the instantaneous star formation rate (including massive stars that quickly die as supernovae) is given by
\begin{equation}
  \frac{{\rm d}\rho_\star}{{\rm d}t} = \frac{\rho_c}{t_\star},  \label{SHeqn1}
\end{equation}
with
\begin{equation}
t_\star(\rho) = t_0^\star \left(\frac{\rho}{\rho_{\rm th}}\right)^{-1/2}  ,  \label{SHeqn2}
\end{equation}
where $t_0^\star$ is the consumption timescale of the cold clouds at the onset of star formation. Note that since the mass fraction in cold clouds, $x$, is always close to unity in the model, the star formation rate follows a Schmidt-law with ${{\rm d}\rho_\star}/ {{\rm d}t} \propto \rho^{3/2}$.

\begin{figure}
  \centering
\resizebox{8.5cm}{!}{\includegraphics{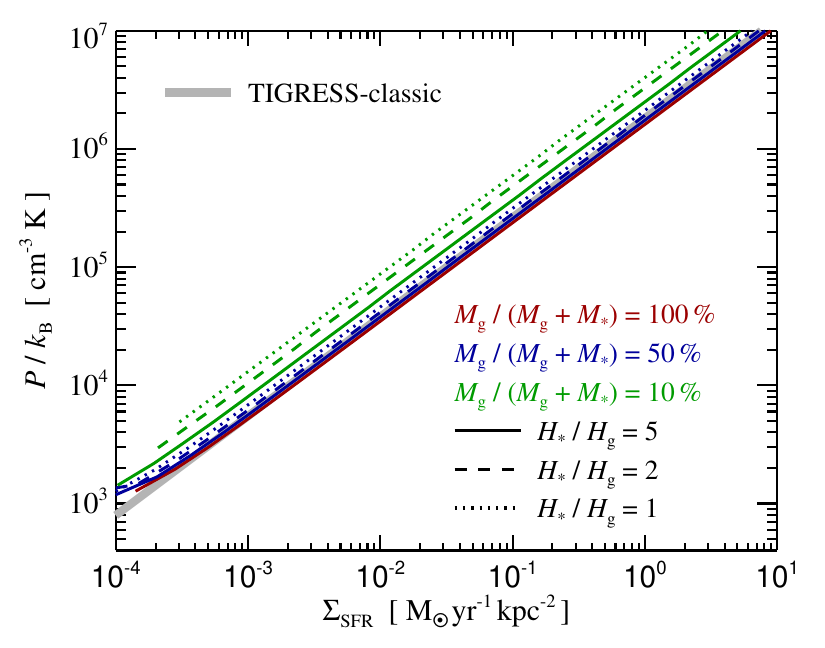}}
\caption{The predicted relation (red solid line) between mid-plane  pressure and star formation surface density for our TIGRESS/Schmidt model where equation~(\ref{prfmsfrlaw}) is adopted as 3D star formation law combined with the equation of state model of equation~(\ref{prfmeqs}). We compare to the power-law fit obtained by \citet{Ostriker2022} for the TIGRESS-classic simulations (grey thick line). We also include results when an additional stellar disk is present, and show cases where the gas disk is equal in mass to the stellar disk (blue), or is substantially lighter so that the gas only makes up 10\% of the total disk mass (green). In both cases, we consider different ratios of the heights of stellar and gas disks, $H_\star / H_g =1$, 2, and 5, as labelled.  Note that this plots shows the mid-plane pressure as a function of prescribed star-formation surface density; the models with different stellar disks exhibit different gas surface densities for a given star formation surface density.\label{Fig:pressure_vs_SFR}}
\end{figure}

SH set the star formation threshold based on the notion that the pressure law should be continuous at the onset of star formation and seamlessly connect at lower density to an isothermal equation of state at a temperature of $10^4\,{\rm K}$, or in other words $P(\rho_{\rm th}) = (\gamma-1)\rho\, u_4$, where $u_4$ corresponds to the thermal energy per unit mass for $10^4\,{\rm K}$ gas, with a mean molecular weight corresponding to a fully ionized mix of hydrogen and helium. SH pointed out that this condition determines $\rho_{\rm th}$ as 
\begin{equation} \rho_{\rm th} = \frac{x_{\rm th}}{(1-x_{\rm th})^2} \frac{\beta u_{\rm SN} - (1-\beta)u_c}{t_0^\star \Lambda(u_{\rm SN} /A_0)},
\end{equation}
with $x_{\rm th} = 1+(A_0+1)(u_c - u_4)/u_{\rm SN}$ being the cloud fraction at the onset of star formation. Here $\beta$ gives the mass fraction of stars that quickly die as supernovae.  SH furthermore showed that the density-dependent cold cloud fraction $x$ can be computed as
\begin{equation}
    x = 1 + \frac{1}{2y} - \sqrt{\frac{1}{y} + \frac{1}{4y^2}} , \label{eqnx}
  \end{equation}
in terms of an auxiliary quantity
\begin{equation}
y = \frac{ t_\star \Lambda(u_h) \rho}{\beta u_{\rm SN} -
  (1-\beta)u_c}.  \label{eqny}
\end{equation}
Here $\Lambda(u)$ is the net cooling function, i.e.~$\Lambda(u) \rho^2$ gives the energy loss rate per unit volume.

\begin{figure}
  \centering
\resizebox{8.5cm}{!}{\includegraphics{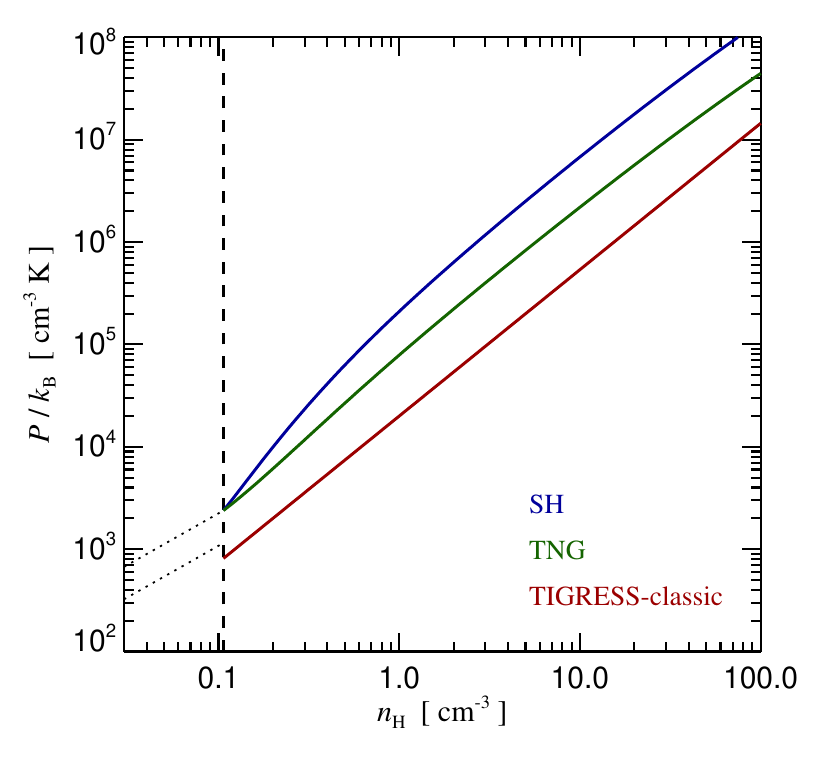}}\\
\resizebox{8.5cm}{!}{\includegraphics{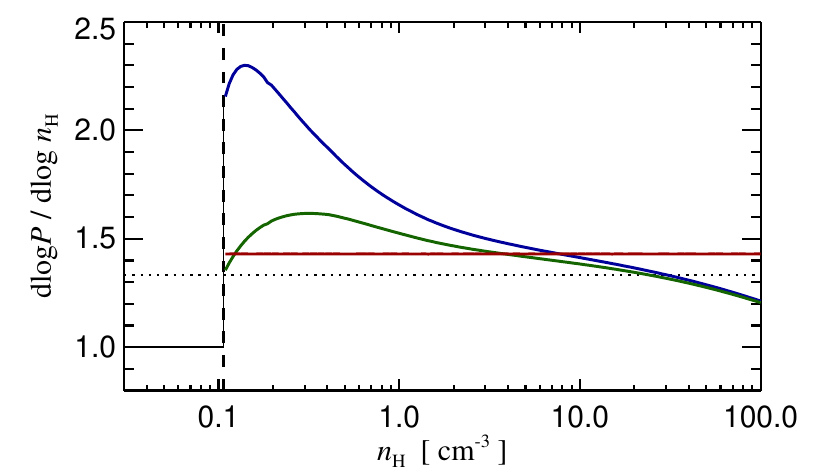}}
  \caption{{\it Top panel:} Pressure as a function of hydrogen number density for three different equation-of-state models, SH, TNG, and TIGRESS-classic, as labelled. The vertical dashed line marks the common star formation threshold density adopted for the models. Below this density, the gas energy equation is solved explicitly, subject to normal radiative cooling and heating. The dotted lines mark the pressure for gas at temperature $10^4\,{\rm K}$ that is either fully ionized (upper line) or fully neutral (lower line). Gas at these densities typically has cooled down to this temperature, with an intermediate ionization state, depending on redshift. {\it Bottom panel:} Logarithmic slope of the different equations of state. The horizontal dotted line marks the critical slope of $4/3$ below which the Jeans mass becomes smaller with higher density. \label{Fig:eos}}
\end{figure}

\begin{figure*}
  \centering
  \resizebox{8.5cm}{!}{\includegraphics{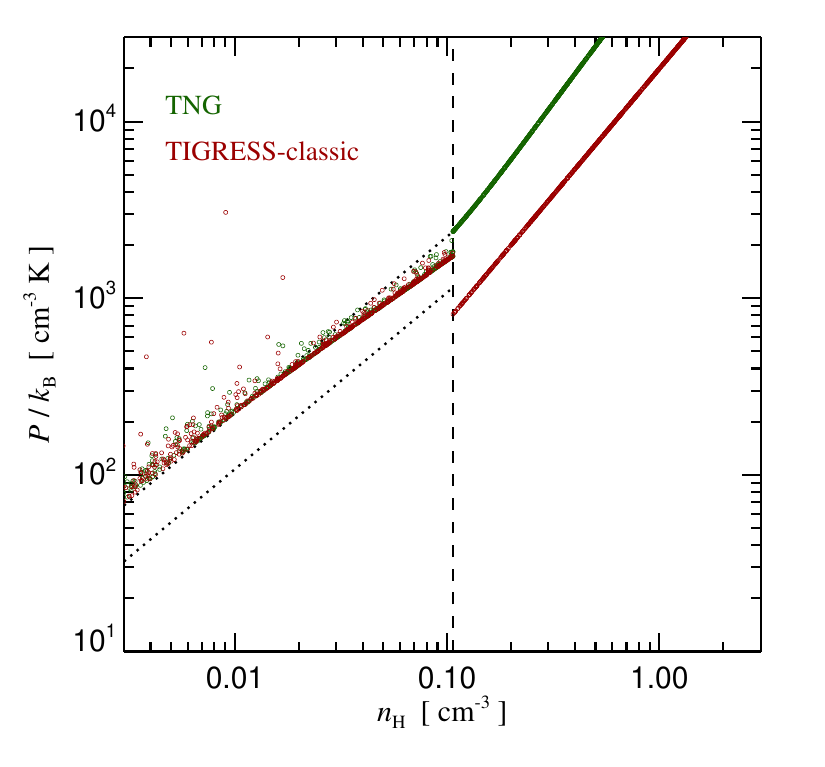}}%
  \resizebox{8.5cm}{!}{\includegraphics{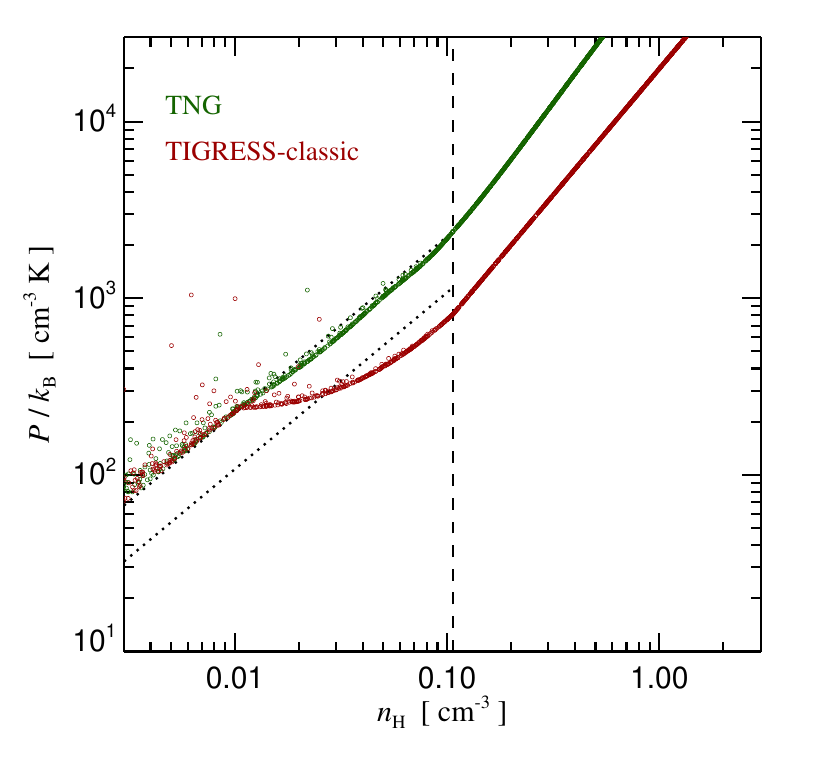}}%
  \caption{Pressure versus density in the regime around the onset of star formation, with individual cells from simulations with TNG (green) and TIGRESS/Schmidt (red) drawn as small circles. At densities above the star formation threshold (dashed vertical line), the pressure follows the equation of state model.  The panel on the left shows the default treatment where the equation of state is  sharply switched on at the star formation threshold, whereas the right panel shows our new variant where the pressure of the EOS model is faded in smoothly below the star formation threshold, so that pressure discontinuities can be avoided. In both cases, the star formation threshold is identical.
    \label{Fig:pressureatthreshold}}
\end{figure*}

\begin{figure}
  \centering
\resizebox{8.5cm}{!}{\includegraphics{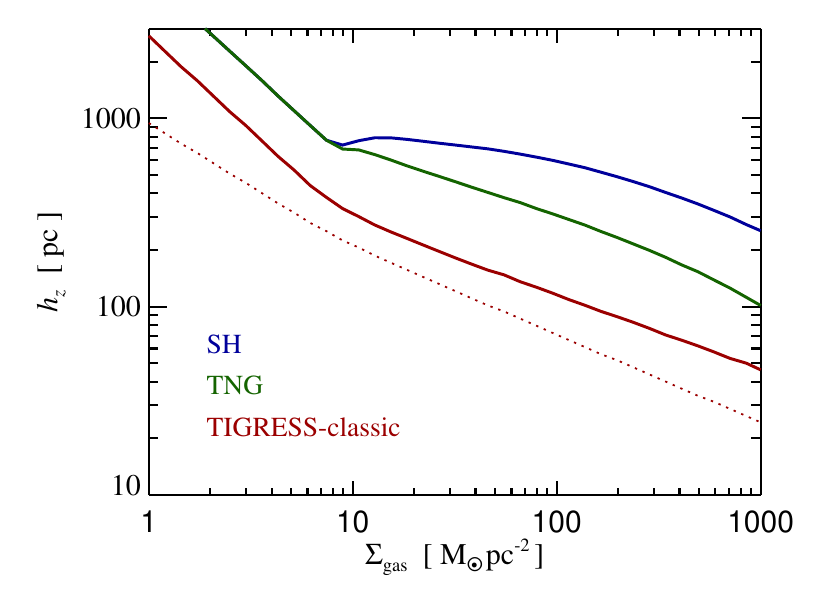}}
  \caption{Expected scale-height of gas layers in plane-parallel symmetry for the Springel \& Hernquist model (blue), the equation of state adopted for IllustrisTNG (green), and the  TIGRESS-classic fit. Below the star formation threshold, an isothermal gas at $10^4\,{\rm K}$ is assumed in all three scenarios. The feedback from star formation leads to a thickening of the gaseous layer beyond the isothermal $h_z\propto \Sigma_{\rm gas}^{-1}$ scaling. The dotted line shows the cell size of TNG50 (with mass resolution $m_{\rm gas} = 8.5\times 10^4\,{\rm M}_\odot$) at the corresponding mid-plane gas densities. \label{Fig:thickness}}
\end{figure}

SH treated $u_{\rm SN}$, $u_c$, $\beta$ and $A_0$ as fixed parameters of the model with physically motivated values, while $t_0^\star$ was treated as a tunable quantity set to approximately reproduce the observational star formation relation of \citet{Kennicutt1998}. Once $t_0^\star$ is determined, it then fully specifies the equation of state law $P_{\rm eff} = P_{\rm eff}(\rho)$ given by equation~(\ref{eqnSH}). This can be seen by noting that one can first compute the star formation threshold $\rho_{\rm th}$ for the parameters at hand. For a given total gas density $\rho$, this then allows a computation of the cold cloud fraction via equations~(\ref{eqny}) and (\ref{eqnx}), while equation (\ref{eqnhot}) gives the temperature of the hot phase, and equations~(\ref{SHeqn1}) and (\ref{SHeqn2}) yield the star formation rate.

We note that in the equilibrium state described by the model, the energy input by supernova is exactly offsetting the radiative cooling losses from the hot gaseous phase, thereby maintaining the finite pressure of the ISM in the presence of rapid cooling. This highlights that already in this simple model there is a tight relation between feedback and the dynamically relevant pressure of the ISM.

\subsection{The IllustrisTNG model}  \label{tgnmodel}

The IllustrisTNG simulation project \citep{Springel2018, Pillepich2018, Marinacci2018, Naiman2018, Nelson2018} is a suite of hydrodynamical cosmological simulations in three different volumes, using a physics model for galaxy formation that incorporates radiative cooling processes, star formation, and supernova feedback \citep{Pillepich2018model}, as well as supermassive black hole growth and associated energy input \citep{Weinberger2017}. The IllustrisTNG (or simply TNG for short) simulations have been quite successful in at least approximately reproducing many galaxy properties and a diverse set of observational data, but some of the assumptions entering the calculations are uncertain or even questionable, motivating the development of improved simulation models that are accounting for modern results for the physics of the ISM.

In particular, the ISM model of IllustrisTNG still relies on the elementary SH model described above, with added heuristic modifications. Specifically, the equation of state model of TNG is taken to be a linear interpolation between the EOS of SH and an isothermal equation of state, viz.
\begin{equation}
  P_{\rm TNG} = q_{\rm EOS} P_{\rm SH}  + (1-q_{\rm
    EOS}) P_{\rm iso} , \label{EOSsoftening}
\end{equation}
where $q_{\rm EOS}$ is a dimensionless parameter that can be used to interpolate between the SH equation of state model (for $ q_{\rm EOS} =1$) and an isothermal equation of state, $P_{\rm iso}(\rho)= (\gamma -1) \rho u_4$, at $T\simeq 10^4\,{\rm K}$ (for $ q_{\rm EOS} =0$).  This approach has been introduced by \citet{Springel2005} due to the realization that the SH model produces a quite stiff EOS that endows disk galaxies with a comparatively thick gas layer and a correspondingly high -- presumably excessively high -- degree of stability that may overly dampen the formation of spiral arms or stellar bars. Given the highly approximate nature of the SH model, a softening of the EOS in the form of equation~(\ref{EOSsoftening}) was suggested as a viable empirical remedy of this problem, and IllustrisTNG selected this approach with a softening parameter of $q_{\rm EOS} =0.3$.

In addition, TNG adopted the parameter choices $\beta =0.226$, $A_0=573$, $T_{\rm SN}= 5.73\times 10^7\,{\rm K}$, and $t_\star = 3.276\,{\rm Gyr}$, slightly different from the values chosen in \citet{Springel2003}. In the following, we will also use these values in all applications of the SH model, so that a primary difference between the SH and TNG parameterizations lies in the softening of the equation of state. For practical purposes, the SH model can be viewed as a special case of the TNG parameterization, but with $q_{\rm EOS}=1$. Aside from this, there are also differences in the star formation law (see below). The star formation threshold is equal, however, and given by $\rho_{\rm th} = 2.516 \times 10^{-3}\, {\rm M}_\odot \,{\rm pc}^{-3}$, corresponding to $n_{\rm H} = X_{\rm H} \rho_{\rm th} / m_{\rm p} \simeq 0.1\,{\rm   cm}^{-3}$ hydrogen atoms per cubic centimeter, for a hydrogen mass fraction $X_{\rm H}=0.76$.

We note that the TNG equation of state eventually gets softer than the critical slope of $4/3$ (i.e.~$\frac{{\rm dlog}\,P}{{\rm dlog}\,\rho} < 4/3$) at which point the Jeans mass becomes smaller with higher density. This is a consequence of the underlying SH model. However, together this can create a numerically quite unfavourable situation -- if a gas cloud happens to be compressed to densities beyond this point, it will start to collapse away, reaching ever higher densities and ever shorter timesteps. But the star formation efficiency -- defined as the depletion time in units of the free fall time, $\epsilon_{\rm ff} = t_{\rm ff} / t_\star$ -- does not become shorter with higher density in the TNG parameterization, rather it is constant with $\epsilon_{\rm ff}$ being substantially smaller than unity. Numerically following the collapse until the gas turns fully into stars then requires several collapse timescales on progressively shorter timesteps, something that becomes computationally rather expensive, especially if a magnetic field is present which can accelerate the drop of the timesteps towards higher densities even further.  

This issue becomes more severe at higher resolution in the TNG model, and was in fact encountered in the IllustrisTNG project for the TNG50 simulation \citep{Pillepich2019, Nelson2019}, where it became so taxing that it prompted a small modification of the TNG model as discussed so far \citep[see section 2.2 of][]{Nelson2019}. This amounted to adopting a more rapid shrinking of the timescale of star formation when the logarithmic slope of the equation of state becomes shallower than $4/3$, which happens at a density $\rho_{\rm rapid} \simeq 230\,\rho_{\rm th}$ for the above parameter choices. For $\rho > \rho_{\rm rapid}$, equation~(\ref{SHeqn2}) was replaced with
\begin{equation}
t_\star(\rho) = t_0^\star \left(\frac{\rho_{\rm rapid}}{\rho_{\rm
      th}}\right)^{-1/2} \left(\frac{\rho}{\rho_{\rm rapid}}\right)^{-1}  ,  \label{SHeqn3}
\end{equation}
effectively steepening the dependence to $t_\star\propto \rho^{-1}$ for very dense gas. This model variation has been the basis of the TNG50 simulation, and we from now on refer to this variant as the TNG model in our subsequent comparisons. We note that tests done at the time \citep{Nelson2019} failed to identify any differences in star formation rates and morphologies of galaxies due to this change, which can be readily understood if only a negligible amount of star formation actually takes place in the simulations at densities $\rho \gg \rho_{\rm rapid}$. Interestingly, however, in massive galaxy clusters with $M_{\rm vir} > 10^{14}\,{\rm M}_\odot$ re-simulated in the TNG-Cluster project, \citet{Nelson2024} found that a side effect of the accelerated star formation in the densest gas is that this can lower the accretion rates of the central supermassive black holes, and so in turn lower the masses of these BHs and their feedback output.

\begin{figure}
  \centering
\resizebox{8.5cm}{!}{\includegraphics{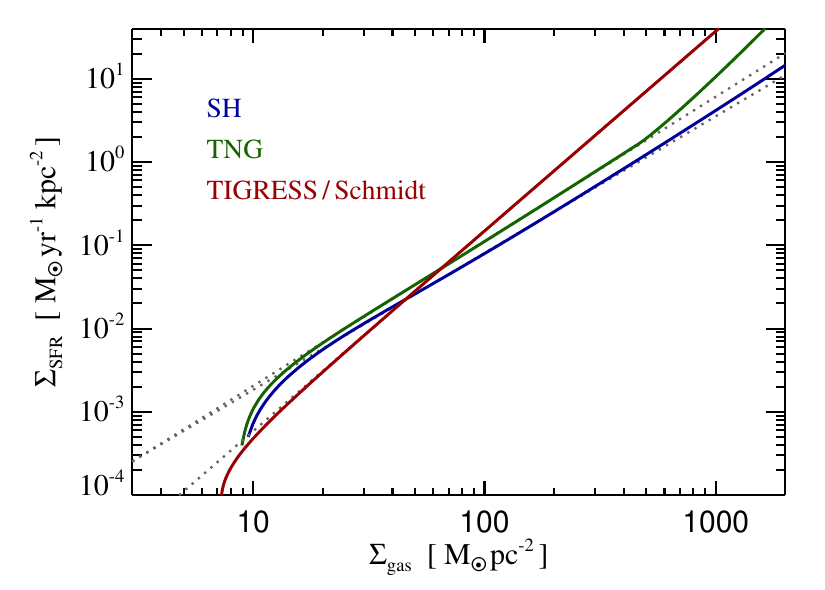}}
\caption{Expected surface density of star formation as a function of gas surface density for the three different EOS models.  The star formation threshold induces a cut-off of the star formation at gas densities around $\Sigma_{\rm gas}\simeq 10\,{\rm M}_\odot {\rm pc}^{-2}$, consistent with observations. The TNG model produces a higher star formation rate density as SH due to its thinner scale height, even though the star formation consumption timescale is the same in SH and TNG, modulo at high density, where TNG adopts an accelerated star formation timescale. TIGRESS/Schmidt on the other has a still softer equation of state, and its star formation timescale depends more steeply on density. As a result, the relation between $\Sigma_{\rm SFR}$ and $\Sigma_{\rm gas}$ is notably steeper. The dotted lines in the background are power-law fits, yielding slopes of 1.64, 1.74 and 2.40 for SH, TNG, and TIGRESS/Schmidt, respectively.
  \label{Fig:sfrvsgas}}
\end{figure}

It is important to note that neither the SH nor the TNG model for the ISM produce galactic winds or outflows by themselves. This is because the multiphase ISM is treated as a single-phase fluid where clouds and hot phase are tightly coupled together in every resolution element by construction, preventing, for example, that hot supernova bubbles can break out of the star-forming layer and vent a hot, low-density wind into the circum-galactic medium. To rectify this problem, SH added a separate,  `by hand' wind-feedback channel to their model, an approach that was likewise adopted by IllustrisTNG, albeit with modified prescriptions for mass-loading and velocity of the winds \citep{Pillepich2018a}. We will in the following first evaluate the models without such an extra ad-hoc wind feedback channel in the isolated disk case, but return to including the wind feedback when running cosmological simulations.

\subsection{Modelling the ISM based on the high-resolution TIGRESS simulations}  \label{SecTigress}

The TIGRESS simulations \citep{Kim2017, Kim2020a, Kim2023} pursue the ambitious goal of resolving star formation in the ISM at sufficiently high resolution and with an accounting of all the relevant physics to predict the feedback regulation of star formation as closely as possible to first principles. To this end, the project models star-forming layers in a shearing-box approximation within a plane-parallel, tall-box geometry. The simulations can therefore be thought of as small pieces excised from a differentially rotating disk galaxy. The modelled physics includes radiative cooling, star formation, heating by stellar radiation, and turbulence driving by supernova explosions. The magnetohydrodynamics is followed with the Cartesian mesh code {\small ATHENA} \citep{Stone2008}.

The theoretical analysis of the TIGRESS \citep{Kim2020a, Kim2020winds}  simulation results by \citet{Ostriker2022} pointed out the central role played by the mid-plane pressure in regulating star formation. Because the pressure is likewise modulated by stellar feedback, this has given rise to the pressure-regulated, feedback-modulated (PRFM) theory of star formation. A basic realization in PRFM is that in vertical dynamic equilibrium the mid-plane pressure must balance the total weight of the gas column above it,
\begin{equation}
P_{\rm tot} = \int_0^{\infty} \rho(z) \frac{{\rm d}\Phi}{{\rm d}z}\, {\rm d}z , 
\end{equation}
which directly follows from equation (\ref{eqnequil}). A foundational conjecture by \citet{Ostriker2022} is that the mid-plane pressure is related to the surface density of star formation, through
\begin{equation}
P_{\rm tot} = \Upsilon_{\rm tot}\Sigma_{\rm SFR}  .  \label{eqnPRFMrel}
\end{equation}
The relation encodes that the effective pressure from turbulence is mainly produced and maintained by feedback related to star formation, although secondary sources from radial transport and gravitational infall can also contribute \citep{Krumholz2016, Krumholz2018}. Note also that the strict linearity of equation~(\ref{eqnPRFMrel}) may be broken by non-linearities in the feedback physics, i.e.~$\Upsilon_{\rm tot}$ is not simply constant. Indeed, \citet{Ostriker2022} find a weak dependence of $\Upsilon_{\rm tot}$ on the star formation rate surface density itself, $\Upsilon_{\rm tot}\propto \Sigma_{\rm   SFR}^{-0.18}$. Direct fits to the time and spatially averaged TIGRESS results yield the power-law relation
\begin{equation}
  \log\left(\frac{P_{\rm tot} / k_{\rm B}}{\rm cm^{-3}K }\right) =
0.840
  \log\left(\frac{\Sigma_{\rm SFR}}{
      {\rm M}_\odot\, {\rm kpc}^{-2\,}{\rm yr}^{-1}}\right) + 6.26  \label{prfmmidP}
\end{equation}
between the total mid-plane pressure and the surface density of star formation.

\begin{figure}
  \centering
\resizebox{8.5cm}{!}{\includegraphics{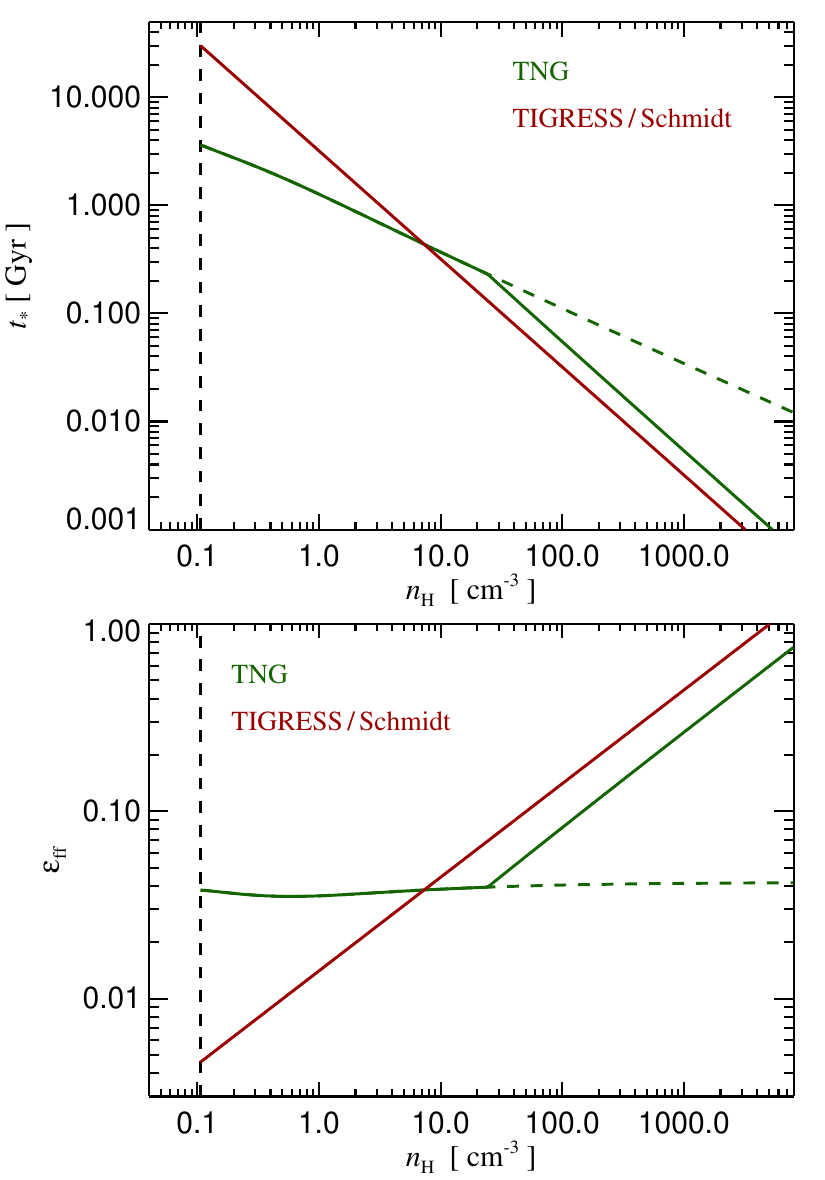}}
\caption{The top panel shows the gas consumption timescale as a function of density for TNG and TIGRESS/Schmidt. The bottom panel expresses this in terms of the efficiency of star formation per free-fall timescale. The dashed line shows the original TNG model before it introduced, for TNG50, an acceleration of star formation at high densities.
\label{Fig:sfrtimescale}}
\end{figure}

The simulations can also be used to measure the relation between mid-plane pressure and total local mass density. While this relation exhibits substantial scatter, \citet{Ostriker2022} show that the means can be well described  by
\begin{equation}
\log\left(\frac{P_{\rm tot} / k_{\rm B}}{\rm cm^{-3}K }\right) = 1.43
\log\left(\frac{n_{\rm H}}{\rm cm^{-3}}\right) + 4.30.  \label{prfmeqs}
\end{equation}
In the following, we adopt this relation as the TIGRESS equation of state, using $n_{\rm H} = \rho X_{\rm H} / m_{\rm p}$ where $X_{\rm H}=0.76$ is the primordial hydrogen mass fraction and $m_{\rm p}$ the proton mass.\footnote{We note that \citet{Ostriker2022} use the conversion $n_{\rm H} = \rho / (1.4\, m_{\rm p})$ more appropriate for Solar abundances, which yields a value that is 6 percent lower, a difference we consider negligible for the purposes of this study.} The corresponding power-law $P(\rho)\propto \rho^{1.43}$ has a slope still a bit steeper than the critical slope $4/3$ needed to stabilize spherical polytropes. Also, for slopes $\gamma > 4/3$, the Jeans mass increases with density. For slopes below this critical value, we expect that local dynamical instabilities  can  potentially break up the disk comparatively easily into gaseous lumps.  Note that the newer TIGRESS-NCR simulations by \citet{Kim2024}, which take metallicity effects into account and explicitly treat UV radiation transfer and photochemistry, suggest a slightly shallower relationship, $P\propto\rho^{1.29}$. Because the relation~(\ref{prfmeqs}) is calibrated against older simulations in the TIGRESS program, we refer to it as the ``TIGRESS-classic'' equation of state. In passing we note that \citet{Jeffreson2024} also find an equation of state substantially softer than TIGRESS-classic, $P\propto \rho^{1.14}$, for their isolated Milky Way-like and NGC300-like simulation models. This emphasizes that the EOS predictions of current high-resolution ISM simulations are still quite sensitive to the exact set of physics that is included, an uncertainty that can hopefully be reduced in the near future.

\subsection{The PRFM model for star formation}

The PRFM theory of \citet{Ostriker2022} does  not directly specify a volumetric 3D star formation law. But a  measurement of the total mid-plane pressure can be  used to infer the star formation rate surface density via relation (\ref{eqnPRFMrel}),  which directly encodes the results of the TIGRESS-classic simulations. Such a pressure measurement is possible provided the vertical structure of both, the gas disk and the stellar disk, can actually be spatially resolved, because only then the mid-plane pressure is trustworthy. However, even in this resolved case, application of the theory in a 3D cosmological simulation code still needs a prescription for distributing the star formation in the dimension transverse to a disk, such that newly created collisionless stellar material can be formed at an appropriate height above the disk. 

\citet{Hassan2024} propose to use the vertically averaged depletion time for this purpose, defined as
\begin{equation}
t_{\rm dep} \equiv \frac{\Sigma_{\rm gas}} {\Sigma_{\rm SFR}}. 
\end{equation}
One can then apply this timescale to every gas cell as
\begin{equation}
\frac{{\rm d} \rho_{\star}} {{\rm d}t} = \frac{\rho}{t_{\rm dep}},
\end{equation}
which yields by construction the intended surface density of star formation, independent of the details of the gas density profile, and independent of additional gravity from a stellar disk or dark matter. Note, however, that the profile of newly created stars is here taken to be proportional to the gas density profile itself, which is an additional assumption that has not yet been verified explicitly against the TIGRESS-classic simulations.

A practical problem in applying this approach in 3D simulations is that one needs to find and define a disk plane on which the central pressure can be measured, which then yields ${\Sigma_{\rm SFR}}$ by means of the TIGRESS-classic results. While this is no problem for isolated disk simulations \citep[e.g.][]{Jeffreson2024}, it is a significant technical challenge in full cosmological simulations and not unambiguously defined for galaxies with morphologies other than disks. In addition, one needs to measure the projected gas density $\Sigma_{\rm gas}$ in order to get the depletion time. Alternatively, one can write the depletion time in terms of a vertically averaged dynamical time \citep{Ostriker2022}, defined as
\begin{equation}
t_{\rm dyn} \equiv \frac{2 H_g}{\sigma_{\rm eff}} = \frac{\Sigma_{\rm gas}}{\sigma_{\rm eff} \rho_0},
\end{equation}
where $H_g \equiv \Sigma_{\rm gas} / [2 \rho(0)]$ is the vertical scale height of the gas, $\rho(0)$ is the gas density in the mid-plane, and
$\sigma_{\rm eff} \equiv [P_{\rm tot} / \rho(0)]^{1/2}$ is the effective sound-speed in the mid-plane. One can then express the depletion time as
\begin{equation}
t_{\rm dep} =  \frac{\Upsilon_{\rm tot}}{\sigma_{\rm eff}} t_{\rm dyn}  = \frac{2 \Upsilon_{\rm tot} H_g}{\sigma_{\rm eff}^2},
\end{equation}
which may allow one to forego a measurement of the projected gas density if instead the gas scale height can be obtained, for example by estimating it from the vertical gradient of the gas density. Since the ISM is expected to be in vertical equilibrium on average, and resolved star-forming ISM simulations show that this is indeed true \citep[see][and references therein]{Ostriker2022}, one may alternatively use the predicted equilibrium value of the dynamical time.  A simple estimate for this in the case that the stellar and gas disks have the same scale height, and dark matter does not significantly compress the disk, is $t_{\rm dyn} = 2/(2 \pi G \rho_{\rm baryon})^{1/2}$, where $\rho_{\rm baryon}$ includes both gas and stars. \citet{Hassan2024} provide more general expressions.

It may also be possible to make the depletion time vary in the vertical direction with the local density by a suitable generalization of the vertically averaged relations, so that defining and measuring 2D-projected quantities on-the-fly during a simulation could be avoided. This is what we consider as the most promising approach to generalize the star formation relations extracted from TIGRESS-classic to applications in (low-resolution) cosmological simulations, something that we will consider in more detail in a forthcoming companion paper. In particular, the PRFM model predicts that both the stellar density and gas density will enter in setting the star formation rate, since the pressure that must be offset by feedback responds to the total gravity.  This can be taken into account via a generalized dynamical time and vertical scale height. In the present study we shall instead first explore a simpler alternative based on conjecturing a 3D star formation law that reproduces relation~(\ref{eqnPRFMrel}) for the case of pure gas disks, and that avoids the need for an explicit identification of the mid-plane pressure, albeit at the price of small deviations from the PRFM prediction when a thin, massive stellar disk is present.  We turn to this model next.

\subsection{A Schmidt law based on the TIGRESS simulation suite}

It is helpful to recall that in equilibrium it is not only in the mid-plane that the pressure balances the weight. Rather, the pressure at any height $z$ needs to balance the weight of the gas column above
it, i.e.~we expect
\begin{equation}
P(z) = \int_z^{\infty} \rho(z) \frac{{\rm d}\Phi}{{\rm d}z}\, {\rm
  d}z  \label{prfmweightprofiles}
\end{equation}
to hold at all $z$. In this sense there is nothing special about the mid-plane, and in fact, \citet{Kim2024} also explicitly show for TIGRESS-NCR that the pressure profiles track the weight profiles (the right hand side of equation~\ref{prfmweightprofiles}) as a function of $z$ \citep[see also][]{Kim2015, Vijayan2020}. We also note that the physics arguments that suggest that the effective pressure should scale with the surface density of star formation also call for it to scale with the 3D density of star formation. In fact, if a purely local relation $\rho_{\rm SFR} = \rho_{\rm SFR}(\rho)$ existed for TIGRESS (like it is assumed in the SH and TNG models), then the equation-of-state~(\ref{prfmeqs}) implies that we should naturally expect a direct relation between star formation rate density and pressure as well.

In the following, we will thus try to parameterize the TIGRESS-classic star formation rate as 
\begin{equation}
 \frac{{\rm d} \rho_{\star}} {{\rm d}t} =
 \frac{\rho}{t^\star_{\rm th}}
\left( \frac{\rho}{\rho_{\rm th}}\right)^{\eta} ,  \label{prfmsfrlaw}
\end{equation}
i.e.~as a power-law
${{\rm d} \rho_{\star}}/ {{\rm d}t} \propto \rho^{1+\eta}$, where the constant $t^\star_{\rm th}$ sets the gas depletion timescale\footnote{Which can equivalently be referred to as gas consumption timescale.} at the onset of  star formation at threshold density $\rho_{\rm th}$. For exploring this conjecture, first note that any given mid-plane density uniquely specifies the central pressure through the equation of state~(\ref{prfmeqs}), and -- if the local stellar density and dark matter density can be neglected -- the gas density profile as well by integrating out the vertical equilibrium structure. Adopting equation~(\ref{prfmsfrlaw}) then also unambiguously determines the surface density of star formation, $\Sigma_{\rm SFR}$.

As we are dealing with power-law relations for the gas component, we can transform equations (\ref{eqnvertical1}) and (\ref{eqnvertical2}) in the absence of stars into dimensionless form for a given mid-plane gas density $\rho_0$ and corresponding central pressure $P_0 = P_{\rm eff}(\rho_0)\propto \rho_0^{\alpha}$ by defining
$\rho = \rho_0 \tilde{\rho}$, $z = z_0 \tilde{z}$,
$q = q_0 \tilde{q}$, with
$q_0 = \left( 4 \pi G \alpha P_0\right)^{1/2}$ and
$z_0 = [ {\alpha P_0} / ( {4 \pi G \rho_0^2)}]^{1/2}$, where
$\tilde{\rho}$, $\tilde{z}$ and $\tilde{q}$ are now dimensionless and fulfil the equations
\begin{align}
  \frac{{\rm d} \tilde{q}}{{\rm d}\tilde{z}}  & =  \tilde{\rho}, \label{eqnvertical3} \\
  \frac{{\rm d} \tilde{\rho}}{{\rm d}{\tilde{z}}} & =  - \tilde{q}
                                                    \tilde{\rho}^{2-\alpha}.  \label{eqnvertical4}
\end{align}
The vertical profile is thus universal under a power-law equation of state, and given by the solution of the two differential equations~(\ref{eqnvertical3}) and (\ref{eqnvertical4}), with the scaling to physical quantities given by $z_0$, $\rho_0$ and $q_0$.

\begin{figure}
  \centering
\resizebox{8.5cm}{!}{\includegraphics{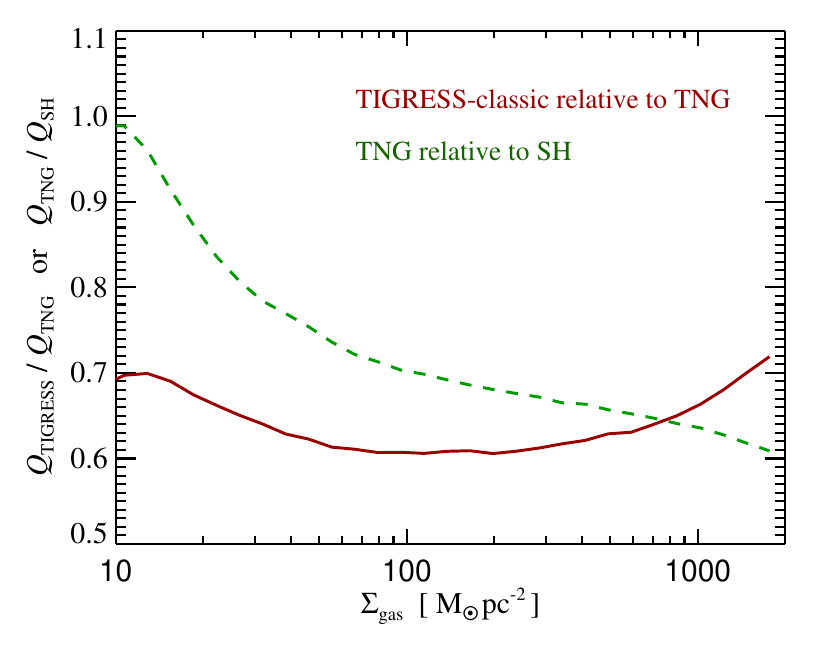}}
\caption{Ratio of the Toomre stability parameter $Q$ between TIGRESS-classic and TNG as a function of the gas surface density of a star-forming disk. Due to the softer equation of state, the effective sound speed in the TIGRESS-classic model is always a factor $\sim 0.6-0.7$ lower than the one in a corresponding TNG simulation, implying a reduction of the Toomre-$Q$ parameter by the same factor. A similar effect is present when going from the correspondingly stiff SH model to the TNG model (dashed line). 
  \label{Fig:toomre}}
\end{figure}

In particular, we expect based on this solution that the height of the gas layer scales as
\begin{equation}
  h_z\propto z_0 \propto \rho_0^{\frac{\alpha}{2} -1},
\end{equation}
 while the  gas surface density should scale as
 \begin{equation}
\Sigma_{\rm gas} \propto \rho_0 z_0 \propto
\rho_0^{\frac{\alpha}{2}}, 
\end{equation}
implying
\begin{equation}
  h_z \propto \Sigma_{\rm gas}^{\frac{\alpha-2}{\alpha}}.
\end{equation}
Furthermore, we expect that the star formation rate surface density scales as
\begin{equation}
\Sigma_{\rm SFR} \propto z_0 \rho_0^{1+\eta} \propto \rho_0^{\alpha/2
  + \eta}
\end{equation}
for the star formation law adopted in equation~(\ref{prfmsfrlaw}). This then implies the scaling relation
\begin{equation}
  P \propto \Sigma_{\rm SFR}^{\frac{2 \alpha}{\alpha + 2 \eta}}  \label{psfrscaling}
\end{equation}
between central pressure and the star formation surface density, and
\begin{equation}
 \Sigma_{\rm SFR}  \propto \Sigma_{\rm gas}^{1 + \frac{2\eta}{\alpha}} 
\end{equation}
for the relation between the surface densities of star formation and gas.

We now recall that \citet{Ostriker2022} have determined $\alpha=1.43$ for the index of the equation of state, and the value 0.84 for the power-law index on the right-hand side of equation~(\ref{psfrscaling}), see equation~(\ref{prfmmidP}). Consistency then determines the value of $\eta$ to be $\eta = 0.987$, which we shall round up to $\eta \simeq 1$ in the following.

We thus conclude that the average properties of the TIGRESS-classic simulations as summarized in Section~\ref{SecTigress} are consistent with a 3D star formation law ${{\rm d} \rho_{\star}}/ {{\rm d}t} \propto \rho^2$ in a situation where the presence of a stellar disk  can be neglected (for example because it is much lighter than the gas disk, or its scale height is very large) and vertical gravity from dark matter can be neglected as well. Incidentally, this is the classic Schmidt law first proposed in his seminal paper \citep{Schmidt1959}. For the relation between scale height and gas surface density we then  expect $h_z \propto \Sigma_{\rm gas}^{-0.40}$, and for that between the surface density of star formation and the gas surface density, $\Sigma_{\rm SFR} = \Sigma_{\rm gas}^{2.40}$.

Finally, to fix the normalization constant $t^{\star}_{\rm PRFM}$ we require that the amplitude of equation (\ref{prfmmidP}) is reproduced. To obtain this value one needs to explicitly solve for the vertical structure. We then obtain $t^{\star}_{\rm th} = 33\, {\rm Gyr}$ where we have adopted for definiteness the same reference density $\rho_{\rm th}$ that we use in the TNG/SH models as star formation threshold. Note that at densities of $n_{\rm H} \simeq 1\, {\rm cm}^{-3}$, the gas consumption timescale for star formation is then about $3.3\,{\rm Gyr}$, while at $n_{\rm H} \simeq10\, {\rm cm}^{-3}$ it already drops to $0.33\,{\rm Gyr}$.

In the remainder of this paper we will study this model as a simplified version of the PRFM theory. We refer to it as ``TIGRESS/Schmidt'' model, reflecting the combination of TIGRESS-based equation of state and star formation law it represents. Note that in this model we do not explicitly use the relation~(\ref{prfmmidP}) between surface density of star formation and mid-plane pressure in the simulations, rather we expect it to emerge from the simulations. In particular, for pure gas disks we would expect this relation to be reproduced quite accurately, whereas in situations with prominent thin stellar disks  deviations may occur which would highlight that this is only an approximate realization of the PRFM theory. 

We now turn to explicitly verifying the above expectations by numerically integrating the star formation rate predicted for the vertical density structure obtained as solution of the differential equations governing the vertical equilibrium. When this is repeated as a function of the central pressure, one obtains the relation between mid-plane pressure and surface density of star formation. We carry out this calculation both for pure gas disks, as well as for cases where an additional stellar disk of certain mass and height relative to the gas disk is added. We parameterize the stellar disk density profile as 
\begin{equation}
\rho_\star(z) = \frac{\Sigma_*}{2 z_0} {\rm sech}^2 \left(\frac{z}{z_0}\right),
\end{equation}
and adjust $z_0$ such that the ratio of gas and stellar vertical scale heights, defined as $H_g = \Sigma_{\rm gas}/[2\rho_g(0)]$ and $H_\star = \Sigma_{\star}/[2\rho_\star(0)]$, has a prescribed value. 

The results are shown in Figure~\ref{Fig:pressure_vs_SFR},  where they are compared to the power-law relation~(\ref{prfmmidP}) obtained from the TIGRESS-classic simulations. As expected based on the analytic considerations above, we obtain very good agreement for pure gas disks, confirming that a description consistent with the PRFM theory is obtained in this case. When including stellar disks, we consider cases where the gas disk makes up 50\% or 10\% of the total disk mass, and for the thickness ratios we consider $H_\star/ H_g = 1$, $2$, or $5$. While for the thick stellar disks the results are hardly affected, the additional gravitational pull of the stellar disk can be felt more strongly for thin and massive stellar disks, and this modifies the vertical equilibrium structure of the gas at fixed mid-plane pressure. Our model then predicts somewhat less star formation at a given mid-plane pressure than forecast by the PRFM theory. 

Since the PRFM theory is focused on the global effects of feedback for maintaining energy balance in the ISM rather than the specific local conditions required for star formation, it does not make any specific prediction for the vertical profile of star formation.  The TIGRESS simulations do show, however, that star formation is more weighted toward the mid-plane than the gas distribution.  This is qualitatively in line with the non-linear dependence $\dot\rho_\star \propto \rho^2$ for TIGRESS/Schmidt or the $\dot\rho_\star \propto \rho^{3/2}$ for SH/TNG.

\begin{figure}
  \centering
\resizebox{8.5cm}{!}{\includegraphics{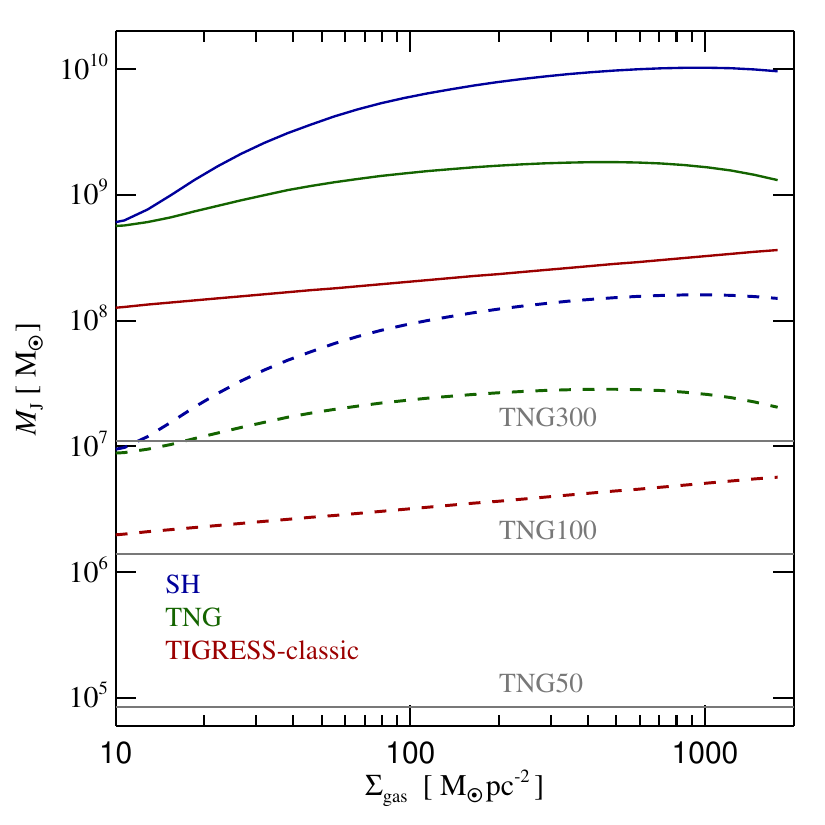}}
\caption{Jeans mass at the mid-plane of star forming disks, as a
  function of their gaseous surface mass density, for three equation
  of state models. For all of them, the Jeans mass reaches a minimum
  at the onset of star formation, because for denser gas the equation
  of state prevents a further decline of the Jeans mass. The
  resolution requirements are thus actually most acute at the onset
  of star formation. According to the \citet{Truelove1997} criterion,
  the Jeans length needs to be resolved with at least 4 resolution
  elements, which translates into mass resolution requirements shown
  with dashed lines. For comparison, the horizontal lines mark the
  baryonic resolutions of the TNG50, TNG100, and TNG300 simulations.
  \label{Fig:jeans}}
\end{figure}

\subsection{Comparison of the three EOS models}

In Figure~\ref{Fig:eos} we show the equations of states  of the three  models SH, TNG, and TIGRESS-classic, assuming the same star formation threshold in all cases. Below this value, the gas is still dense enough to rapidly cool to temperatures of around $10^{4}\,{\rm K}$, but the ionization state of the gas at densities just below the star formation threshold depends on the strength of the UV background. This in turn affects the mean molecular weight, and thus the pressure at this temperature. At redshift $z=0$, we typically find an intermediate ionization state such that neither the assumption of full ionization nor of fully neutral gas would guarantee pressure continuity. A finite pressure jump is however slightly problematic numerically, as it formally implies extremely steep pressure gradients that simply cannot be resolved by the numerical code and will invariably act as a source of numerical noise.

In previous work with the TNG and SH models this problem has been ignored, but here we suggest a simple fix. For densities above $\tilde{\rho}_{\rm th}\equiv \rho_{\rm th}/10$  and below $\rho_{\rm   th}$, we introduce an interpolation parameter
$f = \log ( \rho / \tilde{\rho}_{\rm th} )/  \log ( \rho_{\rm th} / \tilde{\rho}_{\rm th} )$, and define the pressure as
\begin{equation}
P = P_{u}^{1-f} P_{\rm eqs}^f ,  \label{eqnsmth}
 \end{equation}
where $P_{\rm u}$ is the regular pressure computed from the thermal energy per unit mass and the density, and $P_{\rm eqs}$ is the pressure of the equation of state model for the star-forming phase  (if star formation was still allowed at this density). This  prescription thus linearly interpolates in a small density range below the  star forming cut-off between the thermodynamic pressure and the equation of state pressure, so that the latter smoothly fades in at the star formation threshold. In Figure~\ref{Fig:pressureatthreshold} we show the result of this when enlarging  the region around the star formation threshold in realizations of a star-forming disks. In the left panel, we show results for a sudden onset of the equation of state model at the star formation threshold whereas the right panel illustrates the situation when the smoothing prescription of equation~(\ref{eqnsmth}) is adopted. Note that the star-forming phase itself is not affected by this, and neither do we expect significant differences in the star formation rates. However, the smoothed version is numerically better behaved and should be less prone to perturbations triggered by numerical noise and also be better behaved in terms of numerical convergence.

In Figure~\ref{Fig:thickness}, we turn to look at the expected relation between scale height and surface density. Here we see that the gas layers predicted for TIGRESS-classic at high surface densities are substantially thinner than for the other models. At face value this should then also lead to thinner stellar disks, provided other heating effects do not complete wash out the intrinsic thickness at birth.

\subsection{Comparison of the star formation models}

Next, we consider the relation between $\Sigma_{\rm SFR}$ and $\Sigma_{\rm gas}$ predicted for the different equation of state models. This is shown in Figure~\ref{Fig:sfrvsgas} for SH, TNG, and TIGRESS/Schmidt. The slope of the TIGRESS/Schmidt relation is considerably steeper but still lies close to observational inferences \citep{Kennicutt1998, Bigiel2008, Kennicutt2021, Barrera-Ballesteros2021, Sun2023}.

It is also interesting to compare the star formation timescales of the models, and to relate it to the free-fall timescale at a given density,
\begin{equation}
t_{\rm ff} (\rho)\equiv\left(\frac{3\pi}{32 G \rho}\right)^{1/2}   .
\end{equation}
In Figure~\ref{Fig:sfrtimescale}, we show in the top panel the run of the gas consumption timescale, $t_\star = \rho / \rho_{\rm SFR}$, as a function of density, and in the bottom panel the ratio $\epsilon_{\rm ff} =t_{\rm ff}/t_\star$ for our models, which is the efficiency of star formation per free-fall time. For the basic TNG/SH models, this is essentially a constant (dashed line), whereas for TIGRESS/Schmidt this tends to go up towards higher densities. For this reason the TIGRESS/Schmidt model should also be free of the problem encountered for the TNG50 simulation at high density (see the discussion in Section~\ref{tgnmodel}), both because its equation of state stabilizes against fragmentation and never becomes softer than a power-law index of $4/3$, and because of its more favourable scaling of $\epsilon_{\rm ff}$ with density. For TNG50, on the other hand, a steeping of $t_\star \propto \rho^{-1}$ was adopted at an overdensity where its equation of state becomes softer than 4/3, which is indicated as the solid green line in Fig.~\ref{Fig:sfrtimescale}. Incidentally, this is the same as inferred for TIGRESS/Schmidt over the whole density range. At very high densities, the models are therefore quite similar in their star formation rate predictions. Also note that for densities $n_{\rm H} \gtrapprox 1000\,{\rm cm^{-3}}$ the star formation efficiency reaches values close to unity, effectively implying that gas of higher densities is not really expected as gas at such densities should all turn to stars on a single free fall time.

It is also interesting to consider some implications of the different equation of state models for the stability of disks. For self-gravitating, differentially rotating gaseous disks, the \citet{Toomre1964} stability parameter
\begin{equation}
Q= \frac{\kappa\, c_{\rm s}}{\pi  G \Sigma_{\rm gas}}
\end{equation}
can be used to probe for stability against axisymmetric perturbations. The epicycle frequency $\kappa$ does not depend directly on the equation of state but is rather a function of the rotation curve only.  At a given gas surface density, we therefore expect that the ratio of the Toomre-$Q$ for TIGRESS-classic and TNG is simply given by the ratios of their sound speeds, $Q_{\rm TIGRESS} / Q_{\rm TNG} = c_{\rm s}^{\rm TIGRESS} / c_{\rm s}^{\rm   TNG}$, at the mid-plane of the disk. In Figure~\ref{Fig:toomre} we show this ratio as a function of gas surface density, as well as $Q_{\rm SH} / Q_{\rm TNG}$. Other things being equal, we thus expect the TIGRESS model to always exhibit a considerably lower Toomre-$Q$ parameter than the TNG model, which in turn is less stable than SH.

\begin{figure}
  \centering
\resizebox{8.5cm}{!}{\includegraphics{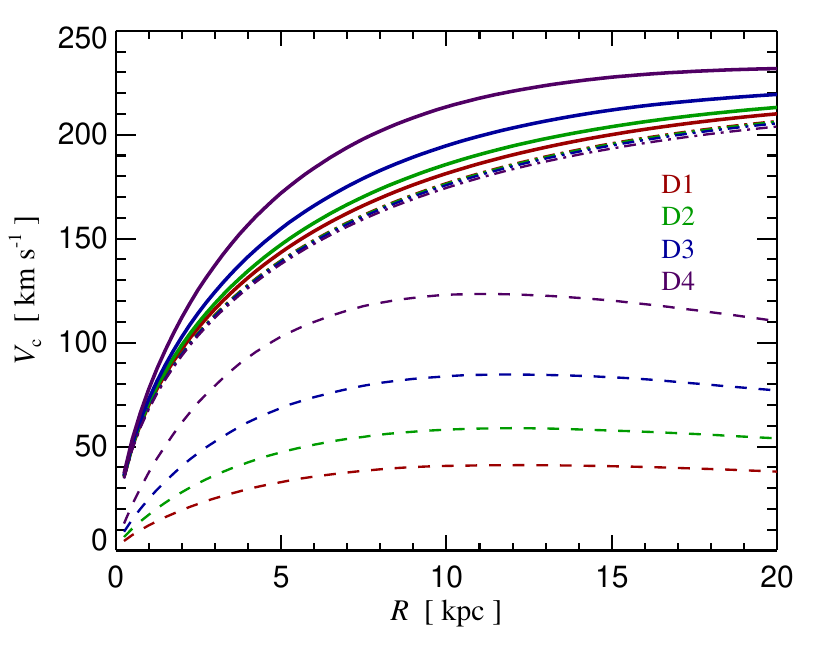}}
\caption{Circular velocity profiles of our isolated disk models. The solid lines show the run of $v_c(R) = [G M(< R) / R]^{1/2}$ where $M(< R)$ is the total enclosed mass inside radius $R$. The dot-dashed lines include only the contribution of the dark matter halo to the rotation curve, whereas the dashed lines are for the gas disks. Our models D1 to D4 increase the gas surface density in steps of factors of two, see Table~\ref{tab:isolateddisks}.
  \label{Fig:vc}}
\end{figure}

\begin{figure}
  \centering
\resizebox{8.5cm}{!}{\includegraphics{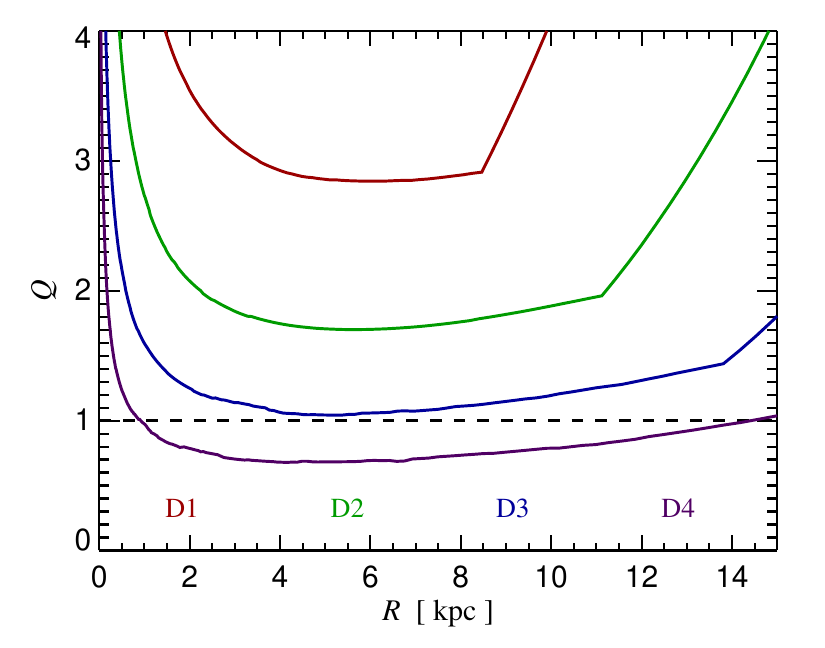}}
\caption{Toomre-$Q$ stability parameter for the gas of our isolated disk models D1 to D4 in the TIGRESS-classic model at the initial time. We see that the high-surface density disk D4 drops below the formal stability boundary of $Q=1$, suggesting that it will likely become unstable after some time of evolution. Note that the TNG and SH models still predict stability even for D4.
  \label{Fig:Q}}
\end{figure}

\begin{figure}
  \centering
\resizebox{8.5cm}{!}{\includegraphics{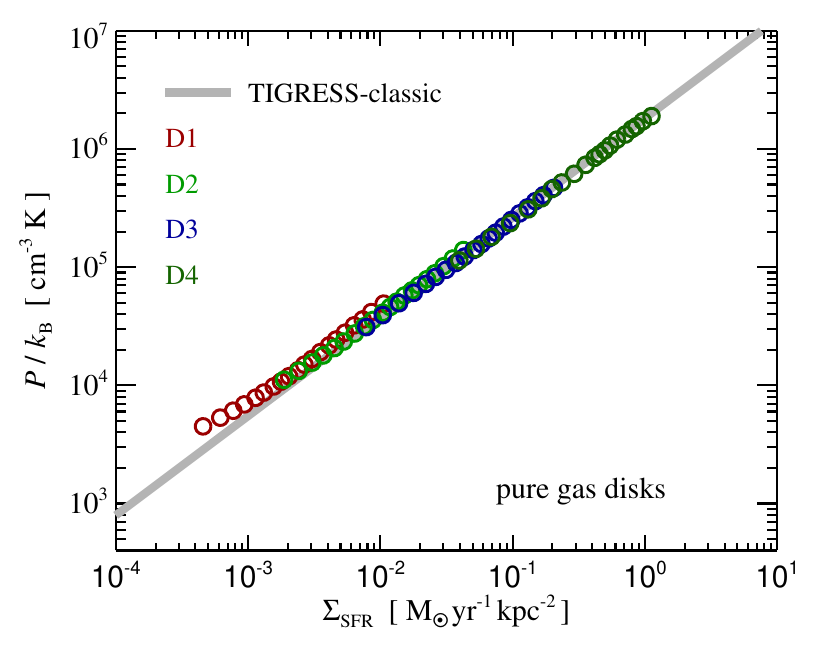}}
\caption{Relation between star formation rate density and central pressure measured in actual realizations of isolated disk galaxies carried out with our TIGRESS/Schmidt model, but with the creation of star particles suppressed so that this can be reproduced over long, quasi-stationary periods of time. Our simulations accurately reproduce the relation measured from the high-resolution TIGRESS-classic simulations.
  \label{Fig:SFRprfmIsolated}}
\end{figure}

\begin{figure*}
  \centering
\resizebox{17.6cm}{!}{\includegraphics{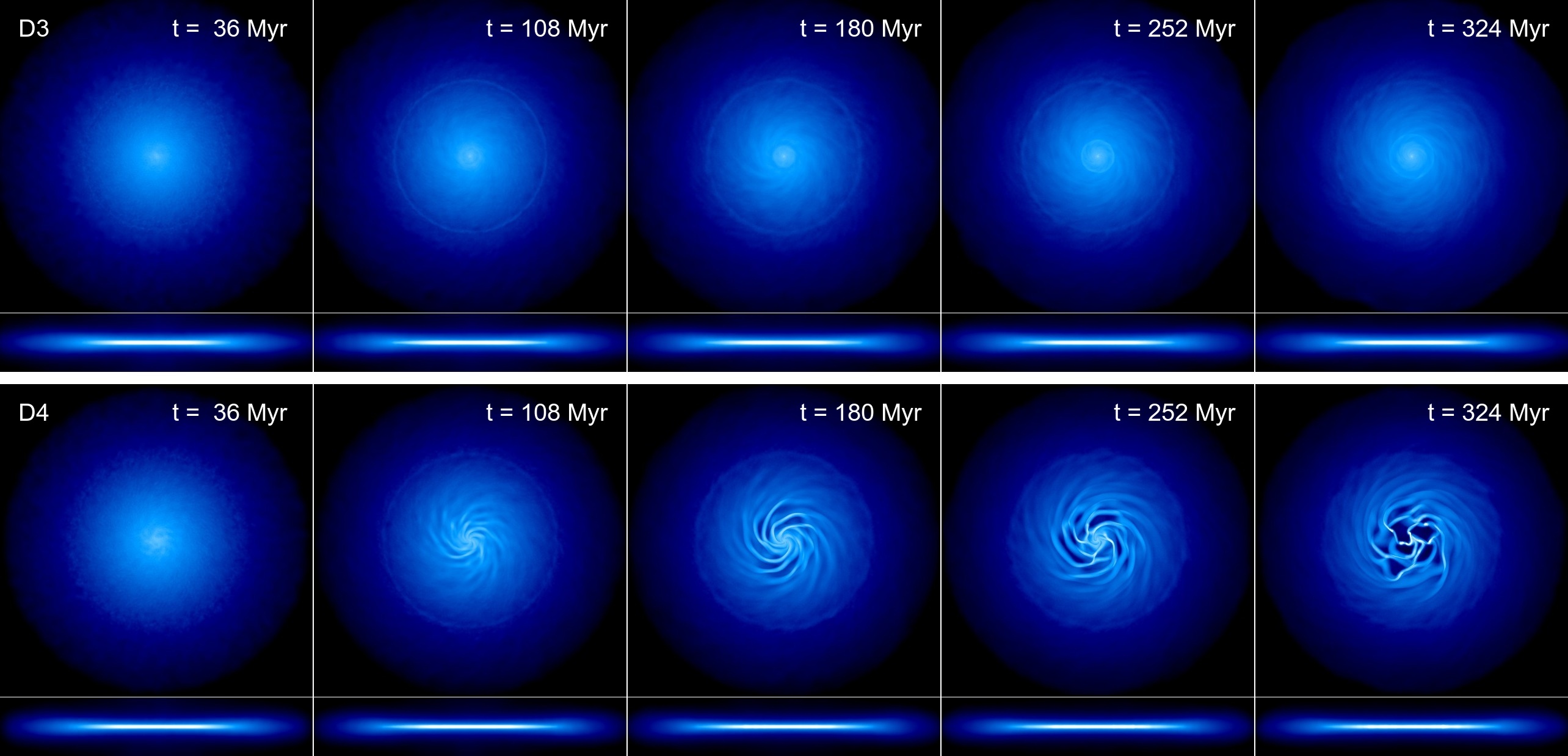}}
\caption{Time evolution of the gas distribution in the isolated galaxies D3 and D4 simulated with the TIGRESS/Schmidt model. It can be seen that there is a stark difference between D3 and D4. Whereas the former remains stable over long timescales, the latter quickly develops instabilities that lead to a break up of the disk into lumps (which is reminiscent of Fig.~6 in SH). This can be understood in terms of the different Toomre-$Q$ stability parameter of the models (see Fig.~\ref{Fig:Q}). 
  \label{Fig:images_gas_isolated}}
\end{figure*}

\subsection{Resolution considerations}

A first indication of the necessary spatial resolution to resolve the vertical scale height of star-forming disks has been shown in Figure~\ref{Fig:thickness}, where we simply included the expected cell size at the central density of the disk for a fixed mass resolution of $m_{\rm gas} = 8.5\times 10^4\,{\rm M}_\odot$ corresponding to TNG50. This cell size needs to be smaller than the vertical scale length to have any chance of resolving the structure.

Another consideration that can be made is to consider the expected Jeans mass of simulations that are regulated by the EOS subgrid models. It is common numerical wisdom that the Jeans length,
\begin{equation}
  \lambda_{\rm J} = \left(\frac{\pi\, c_{\rm s}^2}{G\rho}\right)^{1/2},
\end{equation}
or the Jeans mass $M_{\rm J} = \rho \lambda_{\rm J}^3$ for that matter, need to be resolved by several resolution elements in order to obtain results that are free from spurious numerical fragmentation and thus can at least in principle be trusted \citep[e.g.][]{Truelove1997}. What ``several'' precisely refers to is debated as this also depends on the hydrodynamical discretization method and details of the gravity solver. In Figure~\ref{Fig:jeans} we plot the Jeans mass for the TNG and TIGRESS-classic models. We see that there is a minimum reached at the star formation threshold, simply because the equations of state models provide a slope steeper than 4/3. This is good news in the sense that once the corresponding Jeans mass can be resolved at the onset of star formation, this  will also be the case at higher densities for a Lagrangian code such as {\small AREPO} \citep{Springel2010}, and one does then have a chance for convergent numerical results. Note that for TNG, the Jeans mass eventually starts to become smaller again and would drop below this minimum value at very high densities. However, these  densities are not reached in the model in practice because of the efficient consumption of the gas in star formation already at lower densities.

\section{Isolated disk simulations}  \label{Secdisks}

We set up isolated galaxy models following the method described in \citet{Springel2005}. The compound galaxy models consists of a stationary \citet{Hernquist1990} dark matter potential\footnote{For computational simplicity -- the dark matter halo could also be represented with a live N-body realization, which however would introduce additional numerical noise unless a large particle number is used.}, and a gaseous disk in rotational equilibrium with a surface density declining exponentially with radius. We do not include an initial stellar disk. The size of the disk is related to the assumed angular momentum content of the baryonic material. Our default parameter choices for the models examined here are a virial velocity of $v_{200}=169\,{\rm km\, s^{-1}}$, a spin parameter $\lambda=0.04$, and a dark matter halo concentration of $c=12$. This yields a total halo mass of $1.65\times 10^{12}\,{\rm M}_\odot$ and a rotation curve (see Figure~\ref{Fig:vc}) quite typical for a late-type galaxy without a central bulge.

\begin{table}
\centering
\begin{tabular}{ccc}
\hline
model name & gas disk mass & mass resolution\\
 & $M_{\rm disk}\;[{\rm M}_\odot]$   & $m_{\rm gas}\;[{\rm M}_\odot]$\\
 \hline
 D1  & $6.09 \times 10^9$  & $7.62\times 10^{3}$\\
 D2  & $1.22\times 10^{10}$ & $1.53\times 10^{4}$\\
 D3  & $2.44\times 10^{10}$ & $3.05\times 10^{4}$\\
 D4  & $4.88\times 10^{10}$ & $6.09\times 10^{4}$\\
 \hline
 \end{tabular}
\caption{Initial parameters of our isolated galaxy models in halos of total mass $1.65\times 10^{12}\, {\rm M}_\odot$. At our default resolution, the gas disks are resolved with $8 \times 10^5$ cells.}
\label{tab:isolateddisks}
\end{table}

The vertical structure of the gas is setup in hydrostatic equilibrium following the equation of state models of the three variants considered here. This makes sure that the initial galaxy model is close to equilibrium with only weak initial transients so that the models evolve in a quasi-stationary fashion. We note that we also fill in a low density background at the virial temperature of the halo in order to have a defined gas phase everywhere in our simulation volume which is necessary for a mesh-code as used here. The total mass in this background medium is small enough to not modify the evolution significantly, even though some small amount of it cools down with time onto the gaseous disk.

We have set-up 4 models entitled D1 to D4 that differ in the total mass of the initial gas disk by factors of 2, but are otherwise very similar in structure. Table~\ref{tab:isolateddisks} gives an overview of some of the key parameters of these models, such as the gas disk masses and the employed mass resolution at our default resolution of $8\times 10^5$ cells. In Figure~\ref{Fig:Q} we show the corresponding profiles of the Toomre-$Q$ stability parameter for the four models, adopting the TIGRESS-classic equation of state. According to this we expect the models D1 to D3 to be stable -- but not the high-surface density model D4 -- something we are going to verify explicitly later on.

\subsection{Pure gas disks}

We first consider test simulations where the equation of state is applied and star formation rates are estimated, but the actual creation of new stellar material is disabled in the code, so that the gas is not really depleted. This serves to verify whether our implementation reproduces the analytic expectations we derived for the TIGRESS/Schmidt model.

In Figure~\ref{Fig:SFRprfmIsolated}, we show measurements of the central mid-plane pressure versus the surface density of the star formation rate in our models D1 to D4 after they have evolved for some time. We use a set of 20 logarithmically space radial rings between $1\,{\rm kpc}$ and $6\,{\rm kpc}$ to carry out the measurements. We see that the analytic expectation (see also Fig.~\ref{Fig:pressure_vs_SFR}) is well reproduced in this case. Note that this also demonstrates that it is a reasonably good approximation to treat small regions in the disk as stratified sheets in the vertical direction with respect to estimating the vertical gravitational field.

We now return to the question of stability, which we analyze in Figure~\ref{Fig:images_gas_isolated} in the form of a time sequence of the evolution of the projected gas density in the D3 and D4 models with the TIGRESS-classic equation of state. While D3 is very stable in time, forming a quasi-stationary state (recall that we have disabled the actual creation of new stellar material here) with a differentially rotating disk whose vertical equilibrium is governed by the equation of state, the D4 model develops instabilities after a relatively short time. While the onset of these spiral patterns can be slightly delayed by using a larger number of resolution elements and perhaps an even more careful construction of the initial conditions to minimize the introduction of residual perturbations, we consider this behaviour unavoidable because it reflects a physical not a numerical instability. The disk is simply too cold to prevent the growth of axisymmetric instabilities. The stiffer equation of state of TNG and SH still keep the model D4 stable, however.

\begin{figure}
  \centering
\resizebox{8.5cm}{!}{\includegraphics{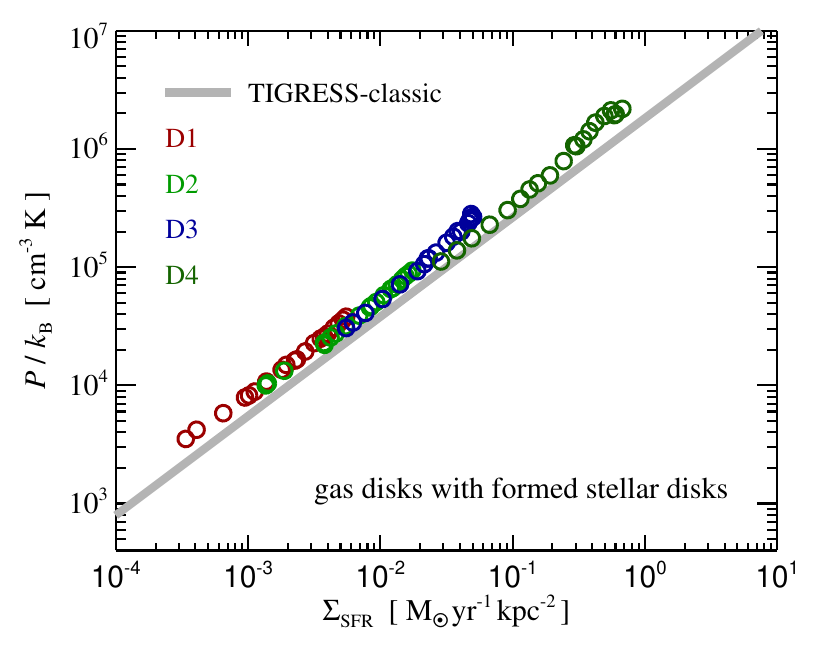}}
\caption{Relation between central mid-plane gas pressure and star formation rate surface density in realizations of isolated disk galaxies with ongoing star formation. The models D1 to D3 are shown 500 Myr after the start of the simulations, whereas the high surface density model D4 is shown after 75 Myr as this model is quite unstable and forms large clumps in the disk afterwards. At these times, up to a bit of more than 20 percent of the disk gas turned into stars. In all cases, we show measurements for 20 azimuthal rings placed onto the disks, logarithmically spaced in radius. The models are computed with our TIGRESS/Schmidt prescription for star formation, and due to the presence of very thin stellar disks (with scale height actually below that of the gas, see Fig.~\ref{Fig:scaleheights_isolated}), deviations from the PRFM theory for this relation are observed, consistent also with our analytic expectations shown in Fig.~\ref{Fig:pressure_vs_SFR}.
  \label{Fig:SFRprfmIsolated_withstars}}
\end{figure}

\begin{figure}
  \centering
\resizebox{8.5cm}{!}{\includegraphics{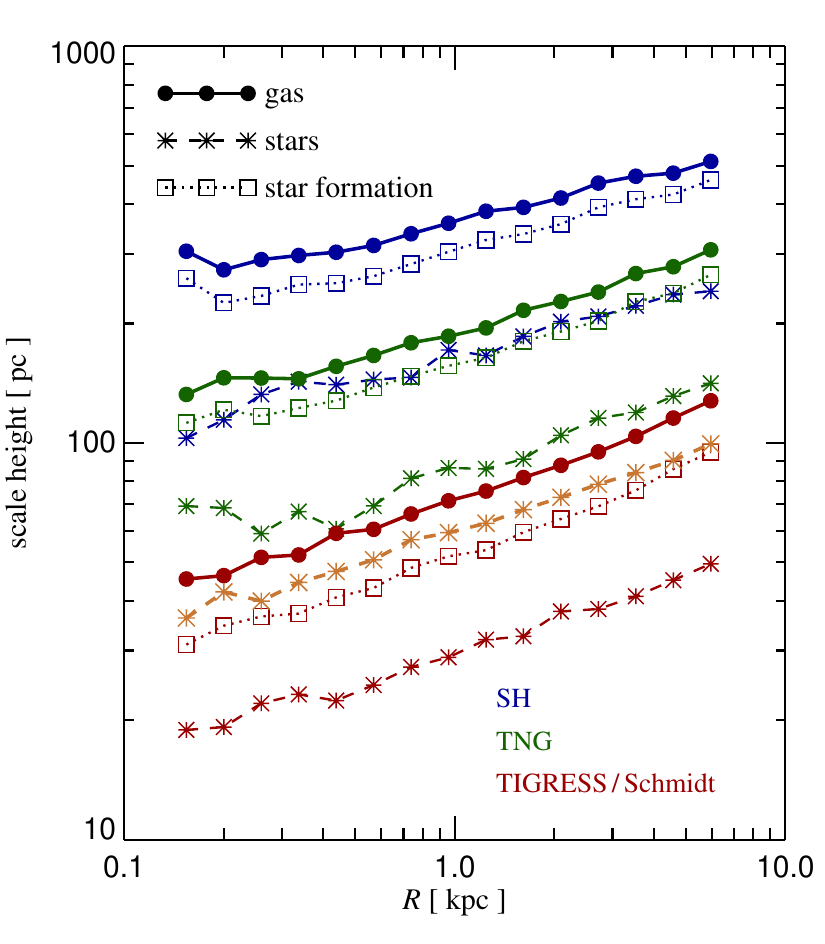}}
\caption{Scale heights of the gaseous disk component (filled points), the star formation distribution (squares), and the formed stellar disk (stars) as a function of radius in our D3 galaxy. We compare results for the three models SH, TNG, and TIGRESS/Schmidt after 400 Myr of evolution. For the TIGRESS/Schmidt simulation, we include an additional scale height measurement (yellow stars) for the stellar distribution, but this time using the coordinates at which the star particles were originally borne.
  \label{Fig:scaleheights_isolated}}
\end{figure}

\subsection{Presence of thin stellar disks}

We now allow the formation of a stellar disk from the ongoing star formation in our isolated disk galaxies, i.e.~the star formation rate is computed and executed as foreseen in the different models.  We are in particular interested in the question whether our Schmidt-like star formation law still approximately reproduces the relation between mid-plane pressure and vertically integrated star formation rate that has been fitted to the TIGRESS-classic simulations.

We show corresponding results in Figure~\ref{Fig:SFRprfmIsolated_withstars}.  Again, we subdivide the disks into 20 bins in logarithmic radius, and then analyse the pressure in the mid-plane of the  gas layer, as well as the corresponding surface density of star formation in the particular radial shell. The results  reproduce the fitting function from TIGRESS-classic quite well, except that they are noticeably shifted above the relation, an effect that becomes slowly stronger with time. This is consistent with the expectations we have outlined in Figure~\ref{Fig:pressure_vs_SFR}, and it can also be viewed as a confirmation of our implementation of the TIGRESS/Schmidt model.

\subsection{The structure of the new stellar component} \label{SecVertStructure}

In Figure~\ref{Fig:scaleheights_isolated} we show the vertical structure of our disk galaxies as a function of radius after they have been star-forming for some time. We select the D3 galaxy for definiteness, and compare the SH, TNG, and TIGRESS/Schmidt models. The other galaxies behave very similarly. We measure as scale heights in this figure the $z$-coordinates that enclose half of the mass of the star-forming gas, half of the stellar distribution, and half of the total star formation rate. This is done separately in 15 radial rings that are spaced logarithmically over the disk of the galaxy. 

Inspection of the results in Fig.~\ref{Fig:scaleheights_isolated} shows that the gas layers are substantially thinner in TIGRESS/Schmidt than in SH, with TNG taking on an intermediate position. This is the expected outcome due to the different pressures delivered at a given gas density by the equation of states of the models, and is consistent with our analytic expectations (Fig.~\ref{Fig:thickness}). Similarly, the radial ``flaring'' reflects the exponential radial decline of the surface density of gas that we imposed in the initial conditions of the galaxy. Note that the star formation has a somewhat smaller scale-height than the gas itself, due to the $\dot \rho_\star \propto \rho^{3/2}$ weighting of the star-formation rate in the SH and TNG models, and the $\dot \rho_\star \propto \rho^{2}$ law in TIGRESS/Schmidt. Because of the steeper dependence on gas density in TIGRESS/Schmidt, the reduction in the star formation scale height is more pronounced in this model, as expected.

What is however perhaps a bit surprising at first sight are the very small scale-heights of the formed stellar distributions, which are much smaller than the scale-heights of the star formation itself. In the current code the star particles inherit their velocities from the gas cells they are born from, which in the case of a gas in vertical and rotational equilibrium is essentially zero, aside from the coherent azimuthal rotational velocity. In other words, the stars are born with nearly vanishing velocity dispersion. In this case the stellar distribution has no adequate vertical support when it is created -- it needs to collapse vertically and find a new vertical equilibrium by virialization. The virial theorem suggests that this can be achieved by a vertical compression by a factor of two, so that the gained binding energy of the stars can be invested into an equal amount of random kinetic motions of the stars. Incidentally, this is the factor we observe in all three models to good accuracy. 

Another case in point is that we can also measure the vertical scale heights of the stars using their birth positions rather than their instantaneous coordinates. Then we obtain the yellow stars included in the figure (just for TIGRESS/Schmidt for visual clarity). This result lies close to the scale-height of the total star-formation, as expected. In fact, it is just slightly above it because of the ongoing consumption of gas which here is accompanied by a slight shrinking of the layer with time, due to the strong gravitational pull of the formed stars hat are highly concentrated towards the mid-plane.

It is clear that the lack of velocity dispersion of the stars at birth, which has been standard practice in most cosmological simulations of galaxy formation thus far, is a point of concern. While for SH and TNG this effect of a vertical narrowing of the stellar distribution was perhaps welcome in order to correct the relatively thick gas layers in these models, the same is not needed in TIGRESS/Schmidt. On the contrary, it appears likely that the extremely thin stellar disks that are formed in this way make the galaxies much more susceptible to perturbations and clump formation than they should. Fortunately, it seems straightforward to come up with a physically motivated solution to this issue. Since the star-forming phase is supported mostly by turbulence if its multiphase structure can be resolved, one could, for example, draw random velocities for stars upon birth that are consistent with the expected turbulent velocities. In fact, in the TIGRESS simulations, the separate contributions from thermal, turbulent, and magnetic contributions to the effective velocity dispersion have been individually calibrated, and these calibrations could be used in assigning initial velocity dispersions to stars at their birth. We shall investigate this in more detail in our forthcoming companion paper.

\begin{figure}
  \centering
\resizebox{8.5cm}{!}{\includegraphics{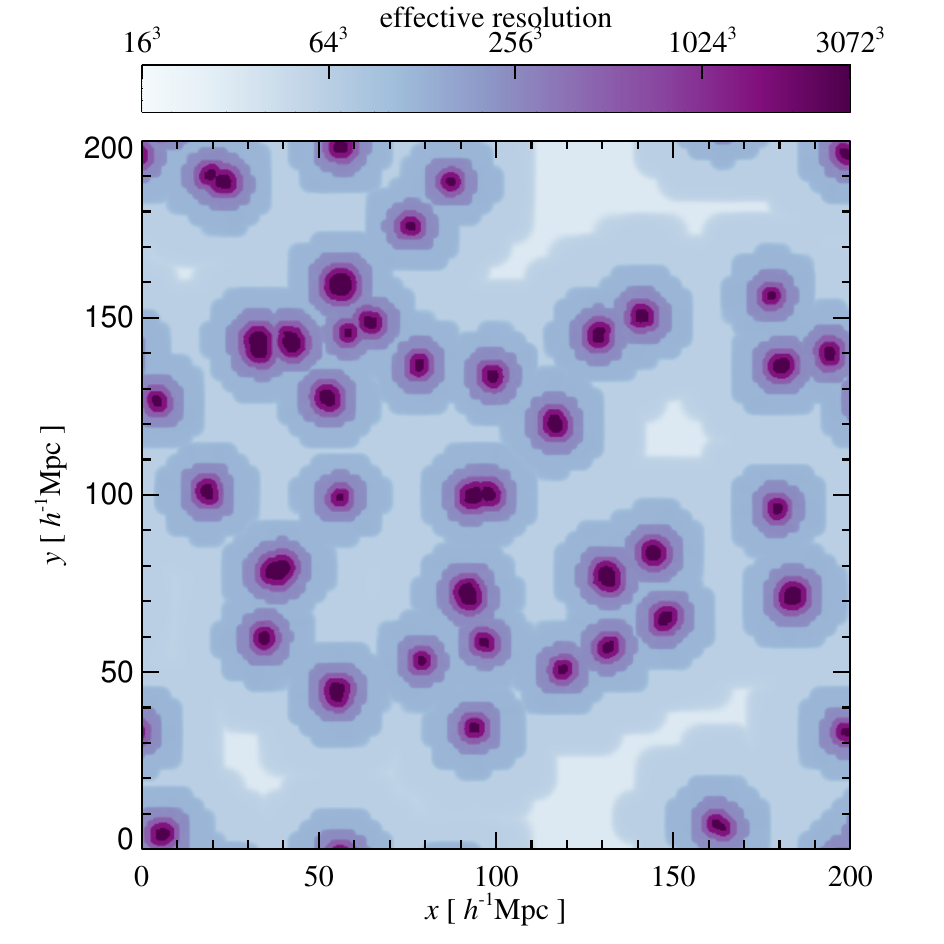}}
\caption{Visualization of a typical Lagrangian particle load of multi zoom-in initial conditions (ICs). Here 42 halos sampling the massive end of the halo mass function have been selected for resimulation from  a $200\,h^{-1}{\rm Mpc}$ parent simulation box. The image shows the effective resolution $(L/d_{\rm mean})^3$ of the particle grid, where $d_{\rm mean}$ is the local mean particle spacing and $L$ is the box size, projected through the whole simulation box. In order to bring out all the high-resolution regions, the projection has been heavily weighted with the effective resolution itself. Note that the highest resolution region at resolution $3072^3$ occupies only a tiny fraction of the volume. The total particle number of the high-resolution multi zoom-in ICs is a factor 5800 lower than a fiducial simulation that follows a homogeneous resolution of $3072^3$.
  \label{Fig:ics}}
\end{figure}

\begin{table*}
\begin{tabular}{cc|c|ccc|c|ccc|c|ccc}
\hline
 &   &  & \multicolumn{3}{c}{$\log (M_{\rm vir} /{\rm M}_\odot) = 11.097 \pm 0.05$} & & \multicolumn{3}{c}{$\log (M_{\rm vir} /{\rm M}_\odot) = 12 \pm 0.05$} & & \multicolumn{3}{c}{$\log (M_{\rm vir} /{\rm M}_\odot) = 12.903 \pm 0.05$} \\
$N_{\rm zf}$ & gas mass resolution & &  \parbox{1cm}{SH\centering} & \parbox{1cm}{TNG\centering} & \parbox{1cm}{TIG/Sdt\centering}  & &  \parbox{1cm}{SH\centering} & \parbox{1cm}{TNG\centering} & \parbox{1cm}{TIG/Sdt\centering}  & &  \parbox{1cm}{SH\centering} & \parbox{1cm}{TNG\centering} & \parbox{1cm}{TIG/Sdt\centering} \\
 \hline
2  &  $3.86 \times 10^6\,{\rm M}_\odot$  &  &   &   &   &&  X & X & X && X & X & X \\
4  &  $4.82 \times 10^5\,{\rm M}_\odot$  &  & X & X & X &&  X & X & X && X & X & X \\
8  &  $6.03 \times 10^4\,{\rm M}_\odot$  &  & X & X & X &&    & X &   &&   &   &   \\
 \hline
 \end{tabular}
\caption{Overview of cosmological `multi zoom-in' simulations carried out with the MillenniumTNG $740\,{\rm Mpc}$ box as parent simulation, using the SH, TNG and TIGRESS/Schmidt (abbreviated as TIG/Sdt) models. We have considered three narrow halo mass ranges of width $0.05\,{\rm dex}$ centred around a virial mass of $10^{12}\,{\rm M}_\odot$, and at an 8 times smaller and 8 times larger mass scale. For each of these three mass ranges, we selected 20 halos randomly among the corresponding objects in the parent simulation and produced zoom-in simulations that refined on these 20 halos. We considered different zoom factors which increase the resolution by factors $N_{\rm zf}^3$, and different models. The crosses mark the models we carried out for this paper.}
\label{tab:zoomsims}
\end{table*}

\begin{figure*}
  \centering
\resizebox{18cm}{!}{\includegraphics{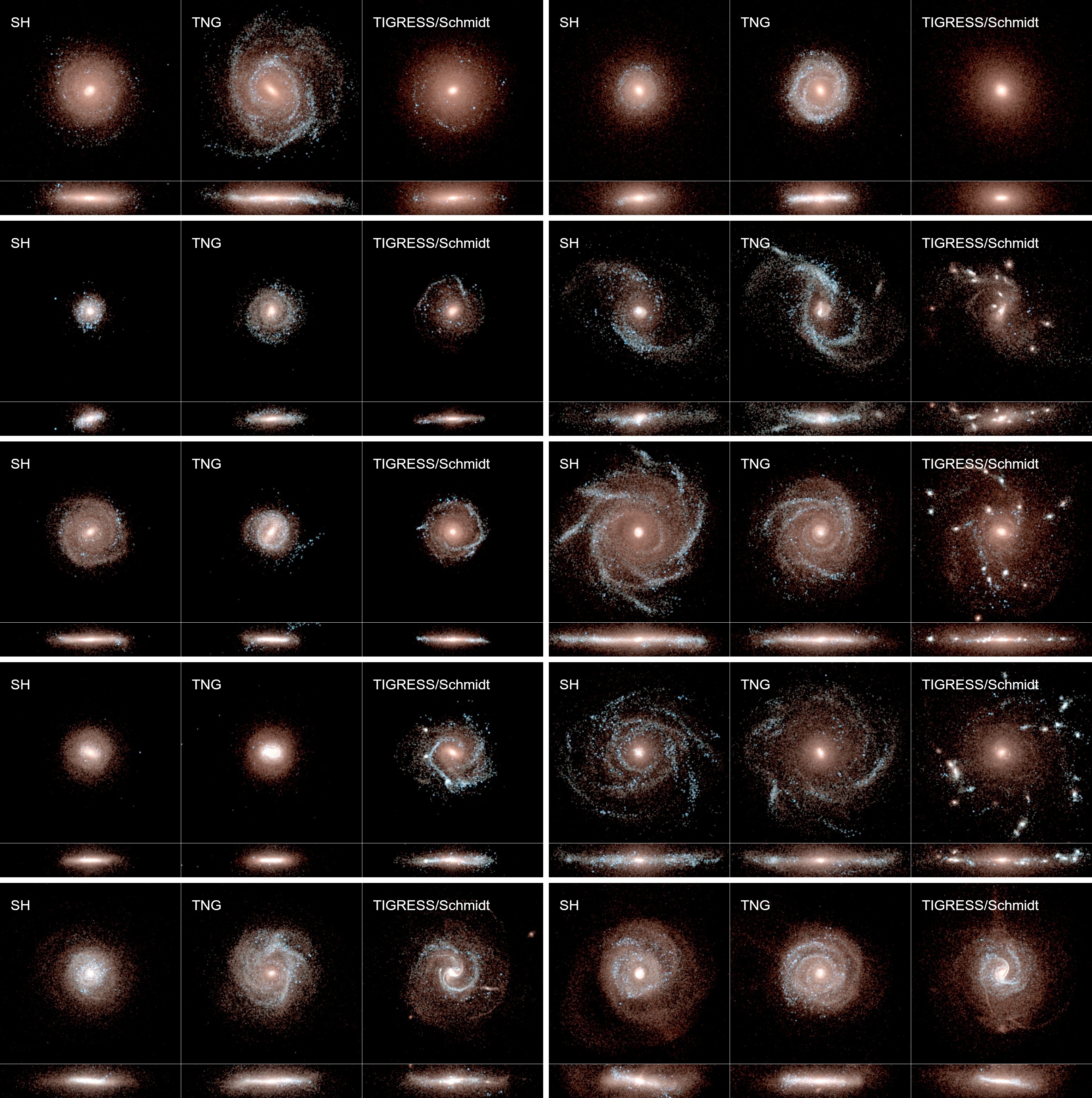}}
\caption{Stellar images of zoom-in galaxies with virial mass around $10^{12}\,{\rm M}_\odot$, simulated with the multi zoom-in technique at zoom factor $N_{\rm zf} =4$ (baryonic mass resolution $4.8 \times 10^5\,{\rm M}_\odot$) as part of a sample of objects randomly selected by mass from the $740\,{\rm Mpc}$ box of the MillenniumTNG project. Each galaxy is simulated three  times, using the SH, TNG, and TIGRESS/Schmidt models. The  systems are shown in face-on and edge-on projections at $z=0$, grouped by the target halo that was selected for resimulation. The individual images show projections of the K-, B-, and U-luminosities of the star particles, mapped to RGB images. All images are $60\,{\rm kpc}$ on a side and use the same logarithmic mapping of luminosity to image brightness and colour. 
  \label{Fig:gals0}}
\end{figure*}

\begin{figure*}
  \centering
\resizebox{18cm}{!}{\includegraphics{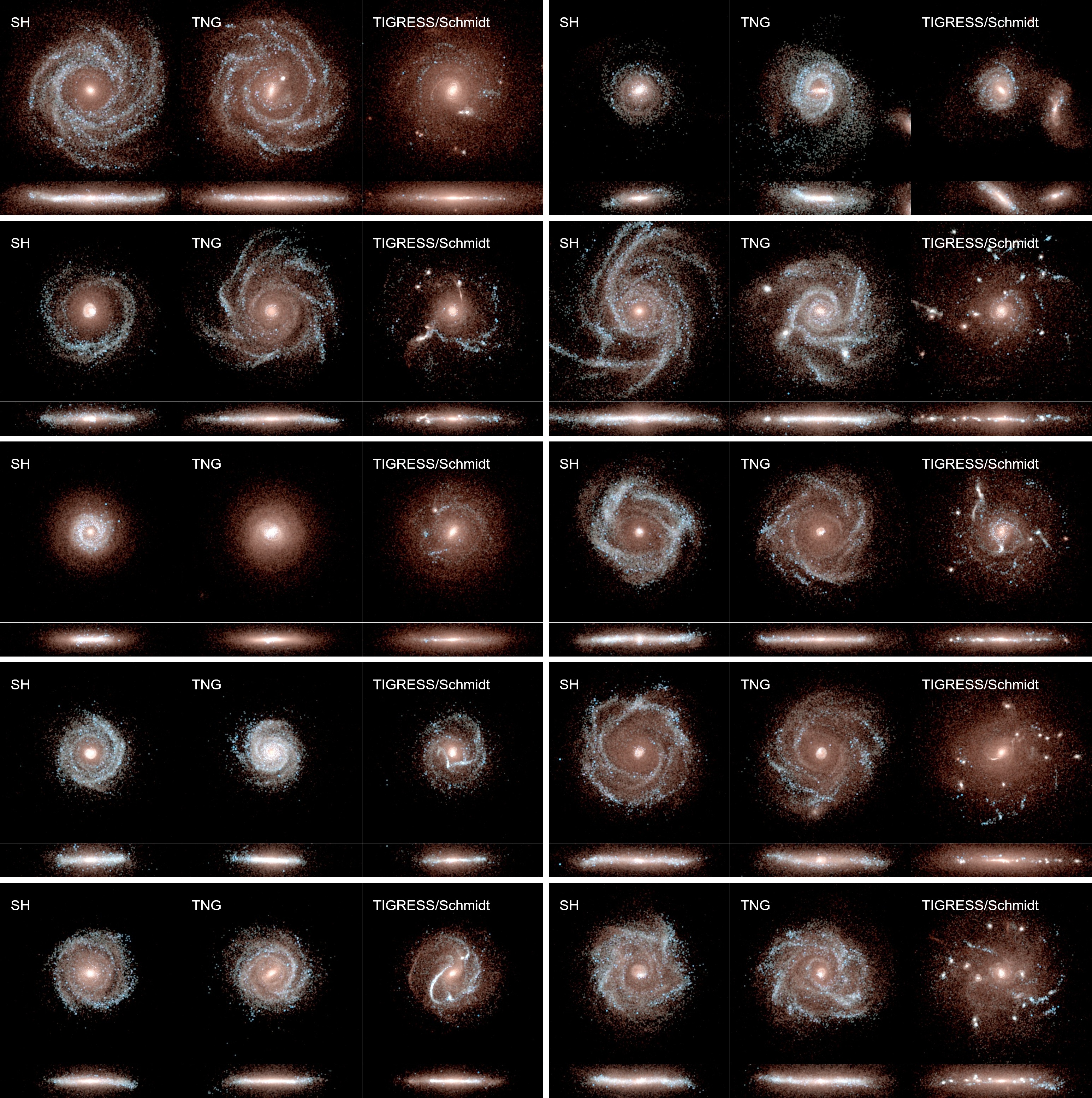}}
\caption{Same as Fig.~\ref{Fig:gals0}, but for the other 10 galaxies from our sample of 20 galaxies selected for the multi zoom-in calculations at halo mass scale $10^{12}\,{\rm M}_\odot$.
  \label{Fig:gals1}}
\end{figure*}

\section{Cosmological zoom simulations}  \label{SecZooms}

In the following, we investigate cosmological zoom-in simulations of galaxies selected from the $740\,{\rm Mpc}$ run of the MillenniumTNG simulation project \citep{HernandezAguayo2023, Pakmor2023}. This calculation has a large enough volume and still a sufficiently good mass resolution to obtain representative samples of halos essentially over the  full virial mass range that supports atomic line cooling. Our strategy is to simulate not just single zooms, but samples of halos at once. This has the advantage to allow a more convenient build-up of representative sets of high-resolution galaxies, without having to evolve the background box multiple times, which makes it computationally more efficient than carrying out a separate zoom-in simulation for each galaxy. An additional benefit is the organizational simplification by being able to carry out a single simulation instead of having to perform a large number individually, i.e.~the laborious (and error prone) shepherding of many simulations can be avoided.

\subsection{Multiple zoom-in regions in a single computational box}

To construct ``multi zoom-in'' initial conditions, we have developed as part of the Learning the Universe collaboration a modified version of the initial conditions module {\small N-GENIC} that is built into the simulation code {\small GADGET-4}. The original version of this IC generator\footnote{N-GENIC was written by \citet{Springel2005a} as part of the Millennium simulation project.} was only able to produce homogeneously sampled cosmological initial conditions in periodic boxes using the Zeldovich approximation. Our new version includes several refinements (such as the use of second-order Lagrangian perturbation theory) and can now optionally accept as input a list of halo numbers that refers to the group catalogue of a previously computed parent simulation output at some target redshift $z_{\rm target}$.  The algorithm then produces zoom-in initial conditions that will reproduce these selected halos at higher resolution when evolved again forward in time, while degrading the resolution in regions of the parent simulation that are far away from the target halos.

A prerequisite for this to work is that the parent simulation's ICs have also been created with {\small N-GENIC}, and that the employed random number seed is kept constant, so that all large-scale modes can be set-up identically in the zoom-in simulation as in the original parent calculation. The list of halos accepted by our code can be composed of just one target halo but may also contain any number of halos. The latter is a novel feature of our approach compared to other zoom-in initial conditions codes that are commonly used. In detail, our method works as follows:

\begin{figure*}
  \centering
\resizebox{15cm}{!}{\includegraphics{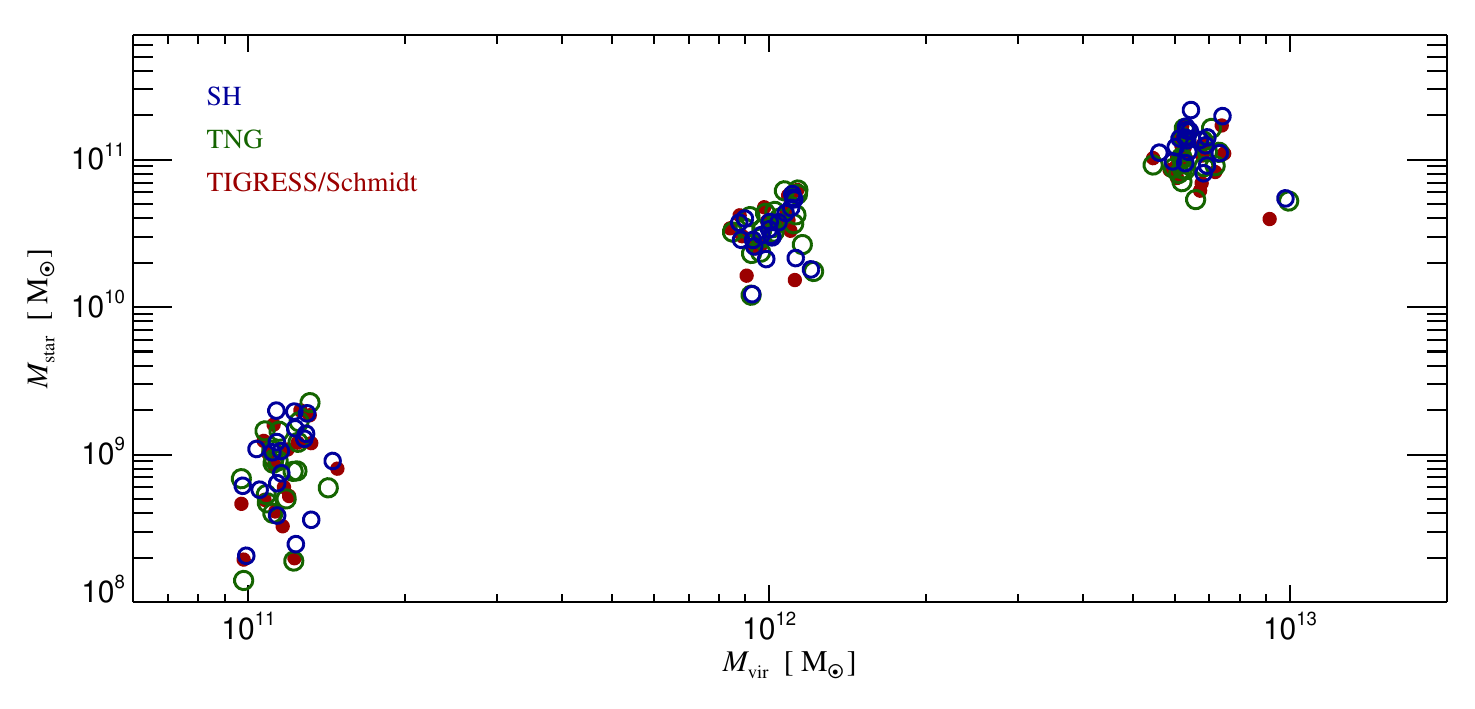}}
\caption{Stellar mass versus virial mass for our full sample of re-simulated halos extracted from the 740 box of the MillenniumTNG simulation suite. We have considered three narrow halo mass ranges of width 0.05 dex around $(1/8)\times 10^{12}\,{\rm M}_\odot$, $10^{12}\,{\rm M}_\odot$ and $8 \times 10^{12}\,{\rm M}_\odot$, and selected 20 objects randomly from them. Each set has then been resimulated with our multi zoom-in initial conditions technique at higher resolution, separately with the SH, TNG and TIGRESS/Schmidt models. We show stellar masses of the central galaxies only, determined by selecting all star particles within a spherical aperture of $25\,{\rm kpc}$ around halo centre. 
  \label{Fig:allstellarmasses}}
\end{figure*}

\begin{enumerate}
  
\item First, all particle-IDs belonging to particles in the selected  halos at the target redshift are loaded and used to determine their Lagrangian coordinates (i.e.~their unperturbed locations at redshift infinity). This is done either by exploiting the fact that the numerical ID-values encode the Lagrangian position directly when {\small N-GENIC} together with an unperturbed Cartesian grid was used for the original parent simulation, or by approximating the Lagrangian coordinates with the particle coordinates assumed in the highest available snapshot output.
  
\item We next use a fiducial grid of size $N_{\rm grid}^3$ to define  the spatial layout of the high resolution regions. Normally $N_{\rm grid}^3$ is chosen equal to the particle number $N_{\rm p}^3$ of the original parent simulation, but this is not a strict requirement. We flag all cells in the grid that contain at least one of the Lagrangian particle coordinates obtained from the previous step.

\item To avoid rough edges of the high-resolution regions marked in this way, and to add a high-resolution buffer region around the virial radius of each halo as a protection against contamination with low resolution material, we smooth and extend the high-resolution regions using a simple neighbours-of-neighbours technique. To this end, we mark in a smoothing step all cells of the fiducial grid as flagged that have at least one neighbouring cell that was already flagged based on a previous step. We repeat this procedure until the total number of flagged cells has grown by a factor $f_{\rm enlarge}$ relative to the initial number; usually we choose an enlargement factor $f_{\rm enlarge} \simeq 2.5$ for this step. The total amount of high-resolution mass will become larger than the summed virial mass of the selected halos roughly by this factor.

\item Next we create a preliminary unperturbed particle distribution by putting down one low resolution particle in every cell which is not flagged (i.e.~lies outside of the high resolution region), whereas all flagged cells are filled with a small grid of particles containing $N_{\rm zf}^3$ particles. We call $N_{\rm zf}$ the `zoom factor'. The mean spacing of these high-resolution particles is chosen as $d_{\rm hr} = L/(N_{\rm zf} N_{\rm grid})$, and they are shifted with respect to the low resolution particles such that the centre-of-mass of the particle(s) representing each grid cell, irrespective of whether they are flagged or not, is preserved.  Note that if the set of high-resolution cells was empty, this step would reproduce the unperturbed particle distribution of the original parent simulation (at least for the standard choice $ N_{\rm grid}=N_{\rm p}$).

\item We now compute a displacement field covering the full box, sampling all modes up to the Nyquist frequency corresponding to $N_{\rm grid}$. Phases and amplitudes of the perturbation modes are uniquely specified just by the random number seed used for {\small N-GENIC}, independent of the specific resolutions chosen, i.e.~matching modes with the parent simulation will be initialized in identical ways. Note also that this step requires an FFT with at least $N_{\rm FFT} \ge N_{\rm grid}$ cells. The resulting displacement field is applied to all low-resolution particles outside the high-resolution region.

\item We next compute a further displacement field covering the full box, which is instead applied to the high-resolution particles only. This extends to the Nyquist frequency of the effective high resolution grid of size $N_{\rm zf} N_{\rm grid}$, provided the available FFT-size is large enough, i.e.~for $N_{\rm FFT} \ge N_{\rm zf} N_{\rm grid}$.  In practice, we may be limited by memory constraints in the maximum Fourier transform size we can carry out, especially if we target large zoom factors and $N_{\rm zf} N_{\rm grid}$ becomes large. So in general we can sample in this step only to the Nyquist frequency of an effective grid given by $N_{\rm eff} = \min(N_{\rm FFT}, N_{\rm zf} N_{\rm grid})$. If we have $N_{\rm FFT} <  N_{\rm zf} N_{\rm grid}$, the minimum is determined by $N_{\rm FFT}$ and extra small-scale power needs to be added in an optional further step, see below.

\item Only if $N_{\rm eff} < N_{\rm zf} N_{\rm grid}$, we compute a further displacement field that is periodic on a scale $L_{\rm box}/f_{\rm fold}$, with a suitably chosen integer folding factor $f_{\rm fold}$.  We need to chose $f_{\rm fold}$ such that $N_{\rm FFT} \times f_{\rm fold} \ge N_{\rm zf} N_{\rm grid}$, hence $f_{\rm fold} =\lceil N_{\rm zf} N_{\rm grid} / N_{\rm FFT} \rceil$. We then sample additional Fourier modes between the Nyquist frequencies corresponding to the $N_{\rm eff}$ and $N_{\rm zf} N_{\rm grid}$ grids. The resulting displacement field (which is periodically replicated to cover the full box) is then added to the displacements obtained for the high resolution particles in the previous step.

\item Finally we downgrade the resolution in the low-resolution volume progressively with distance to the high-resolution regions, with the goal to avoid a modification of the formation history of the high-resolution target halos. We do this by repurposing the gravitational tree-algorithm available in the {\small GADGET-4} code. To this end we first construct an oct-tree for the preliminary particle distribution. For a chosen opening angle $\theta$ we then effectively walk the tree for all high resolution particles by opening all nodes that are seen under a geometric angle larger than $\theta$. Finally, we replace the particle content of all tree nodes that have not been opened by any of the high resolution particles with a more massive particle at the centre-of-mass coordinate and the centre-of-mass velocity of the node. This drastically reduces the number of low-resolution particles, while the gravitational forces for the high-resolution particles stay invariant, at least at the initial timestep and if only monopole forces and the same opening angle are used. Of course, in practice, exact invariance of the evolution of the zoom-in region will quickly be lost, but our distance dependent centre-of-mass coarsening should retain the growth of large-scale structure at the level relevant for influencing the high-resolution patches. Also note that the method does not need to impose geometric restrictions on the permissible shapes, sizes and placements of the high-resolution patches.
  
\end{enumerate}

In the limit of $\theta\to 0 $, the above procedure retains the background resolution fully, and for $f_{\rm enlarge} \gg 1$, any contamination of the high-resolution regions with low-resolution material can be avoided.  The quality of zoom-in initial conditions -- in the sense of computational efficiency versus faithfulness with which the zoom-in halos are reproduced cleanly at high resolution -- is thus mainly a function of these two parameters. We have not yet systematically investigated optimum choices for these parameters and leave this to forthcoming work. Instead, in the following we report on results obtained with the choices $f_{\rm enlarge} =2.5$ and $\theta = 0.5$, which seem reasonably conservative.

In Figure~\ref{Fig:ics} we illustrate our zoom-in initial conditions technique for a simple example of a set of 42 halos that are  sampled over a wide mass range from the halo mass function of a $\Lambda$CDM box of size $200\,h^{-1}{\rm Mpc}$. The map shows a projection of the maximum effective resolution reached along the line of sight, so that the high resolution regions become visible like cell structures in the agar plate of a Petri dish. Note however that the projection hides the smallness of the volume fraction really covered by the high-resolution regions. In this example, $\approx 5 \times 10^6$ high resolution particles are used, whereas doing the full box at the same resolution would require $3072^3 \simeq 2.9\times 10^{10}$ particles, a factor 5800 more. Getting results just for the 42 halos is thus about a 1000 times cheaper computationally with this technique than doing the full box at high resolution.

\subsection{Zoom-in halo samples}

We make use of the multi zoom-in technique to get a reliable first assessment of the differences between the SH, TNG, and TIGRESS/Schmidt effective models  when applied in cosmological simulations of galaxy formation. To this end we consider three sets of halos selected in a large cosmological volume, which for definiteness we identify with the MTNG-740-4320-A model \citep{HernandezAguayo2023} of the MillenniumTNG project (MTNG for short). This calculation used the same cosmological model as IllustrisTNG, but employed a volume  substantially larger (by nearly a factor 15) than TNG300 of IllustrisTNG. The base resolution of the parent MTNG simulation is $4320^3$, yielding a mass resolution close to but slightly worse than TNG300. We keep this as a our fiducial background resolution and consider certain zoom factors relative to it. In particular, $N_{\rm zf}=4$ will give something in between TNG100 and TNG50, while $N_{\rm zf}=8$ has a factor 1.4 better mass resolution than TNG50.

We identify one target halo sample around $10^{12}\,{\rm M}_\odot$ by randomly picking 20 halos at $z=0$ within 0.05 dex around this virial mass value, defined here as the mass enclosing a spherical overdensity of 200 relative to the critical density. Additionally, we pick one halo sample a factor of 8 higher in virial mass, and a further one a factor of 8 lower in virial mass, using a corresponding selection strategy. Table~\ref{tab:zoomsims} gives an overview of the different halo samples and their corresponding gas mass resolutions.  Note that most of the star formation in the universe is expected to happen in halos of virial mass around $10^{12}\,{\rm M}_\odot$ \citep{Behroozi2013}, so this provides an overview of the halos that can be considered the most crucial for building up the stellar mass density in today's universe. We have then  simulated each of the halo samples with a separate multi zoom-in simulation, using either the SH, TNG, or TIGRESS/Schmidt models, at different zoom factors, see Table~\ref{tab:zoomsims}. Ideally, the virial regions of the resimulated zoomed-in target halos should be completely free of any resolution element originating from outside the designated high-resolution patch. This is the case to good accuracy for our halo samples. For the massive halo set, 14 halos have zero contamination, while in 6 halos a very small number of low resolution boundary particles have crossed the virial radius, amounting to less than a fraction of $10^{-4}$ of the halo virial mass in all cases. This low level of contamination tends to grow slightly towards the smaller halo mass samples (reflecting also their, on average, earlier formation time). In our low mass sample, only half of the halos are completely free of contamination, the others show contamination between $10^{-3}$ and $10^{-2}$ of the virial mass, which we still consider negligible for the purposes here. A further reduction of contamination could be achieved by increasing the parameter $f_{\rm enlarge}$ when the ICs are generated. Finally, we note that in principle samples at the same zoom factor could have been combined in a single simulation, but because different zoom factors cannot be mixed at present in {\small N-GENIC} and {\small AREPO} we have refrained from doing this here. Note that for all these simulations we employed the IllustrisTNG model for galactic winds, and we included its models for supermassive black hole growth and feedback.

\subsection{Morphological differences of the simulated galaxies}

In Figures~\ref{Fig:gals0} and \ref{Fig:gals1} we show images of all the 20 galaxies in our $10^{12}\,{\rm M}_\odot$ sample, simulated with the SH, TNG, or TIGRESS/Schmidt models, at zoom factor $N_{\rm zf}=4$. In each case, we show projections of the stellar light in a face-on and edge-on orientation, with the normal direction identified by diagonalizing the moment of inertia tensor of the stellar light. Galaxies at the centres of the same re-simulated halo are matched and shown together side by side. All the images have the same physical size of $60\,{\rm kpc}$  on a side, and the luminosity in the K-, B-, and U-bands is mapped in the same way to an RGB composite image, so that differences in brightness between the images are quantitatively meaningful.

\begin{figure}
  \centering
\resizebox{8.5cm}{!}{\includegraphics{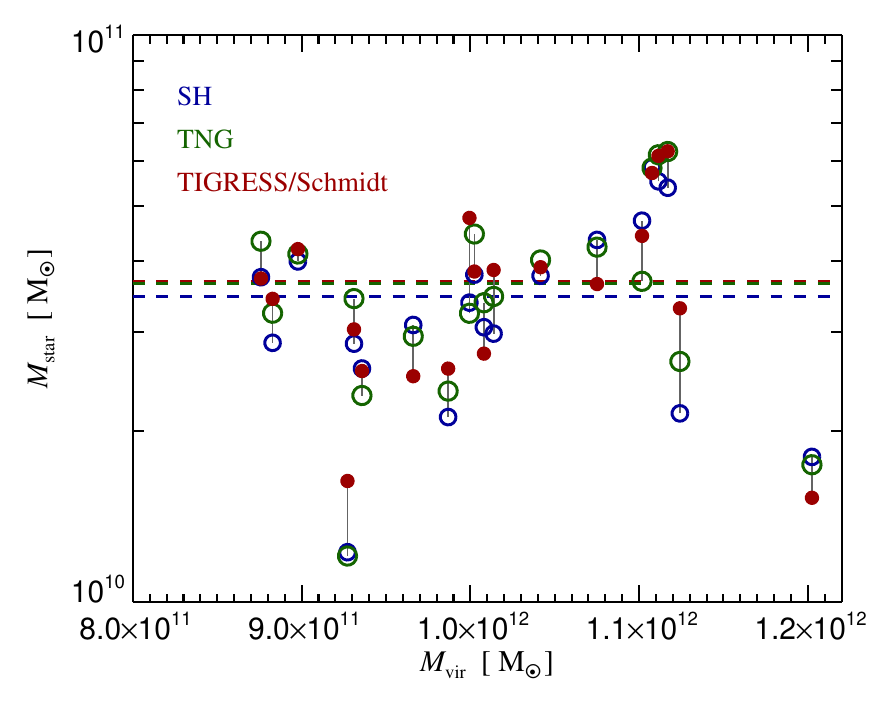}}
\caption{Detailed stellar mass comparison of our $10^{12}\,{\rm M}_\odot$ sample of galaxies, for the SH, TNG and TIGRESS/Schmidt models, as labelled. Corresponding galaxies that are centrals in the same dark matter halos are lined up by thin vertical lines. The very small differences in the virial mass are suppressed in this figure for visual clarity. Horizontal dashed lines show the means of the corresponding samples.
  \label{Fig:stellarmasses}}
\end{figure}

A few general observations can be readily made from these images. While in some cases, significant variations in galaxy size and morphology between the models occur, the overall shape of the galaxies is relatively robust and there is no clearly apparent systematic difference in stellar mass or size between the galaxies. The most striking difference is that the TNG model, and especially the TIGRESS/Schmidt model, show more small-scale structure in the stellar distributions in the majority of objects. In particular, TIGRESS/Schmidt sports in some galaxies marked stellar clumps, hinting that the corresponding evolution went through phases of strong dynamical disk instability related to high surface densities of star-forming gas. This is of course broadly consistent with expectations derived from our earlier analysis of isolated galaxies, and even though the Jeans mass criterion for the TIGRESS/Schmidt model is most stringent, the resolution at $N_{\rm zf}=4$  should be sufficient to suppress numerical fragmentation, so that this fragmentation is likely physical. Whether or not this is a  problem for TIGRESS/Schmidt is less clear however. While some clumpy disk galaxies are observed, recent high redshift observations \citep[e.g.][]{Ferreira2023, Robertson2023} suggest that smooth disks are much more prevalent even at early times than previously thought. But also note that we identified in Section~\ref{SecVertStructure} a potentially quite important shortcoming of the presently employed approach to create collisionless star particles in cosmological simulations. Because they are launched as a dynamically extremely cold distribution without intrinsic velocity dispersion, the resulting stellar disks will be super thin in TIGRESS/Schmidt, amplifying the susceptibility to disk instability. Changing this aspect will reduce clump formation in TIGRESS/Schmidt, something we plan to investigate in forthcoming work.

\subsection{Stellar mass comparison}

We next follow up on this visual inspection with an examination of the stellar masses of the simulated galaxies. To this end we show in Figure~\ref{Fig:allstellarmasses} the distribution of stellar masses as a function of their host halo masses, including results for the three mass bins and the  three star formation models. As expected, we find three narrow clouds in virial mass, reflecting the narrow mass ranges that we selected for the target halos. The virial masses align very closely with the virial masses in the parent dark matter-only simulation, as intended (and even more accurately when hydrodynamics in the zoom-in simulations is disabled). There is a small reduction in the halo mass compared to the dark matter parent simulation, especially for the $\sim 10^{12.9}\,{\rm M}_\odot$ sample, due to AGN feedback, as expected for the TNG physics model. This can be seen as a  further validation of the zoom-in technique. 

Perhaps surprisingly, the results in Figure~\ref{Fig:allstellarmasses} show very little systematic difference -- if any -- in the stellar masses between the SH, TNG, and TIGRESS/Schmidt models, despite the different equations of state. This is true for all three mass ranges considered, and it hints that the stellar masses are fairly insensitive to the details of the EOS treatment.

This can be further corroborated by looking at matched comparisons between the galaxy models in a more detailed fashion. In Figure~\ref{Fig:stellarmasses}, we focus on the $10^{12}\,{\rm M}_\odot$ sample for conciseness, and match the 20 zoomed objects, plotting again their stellar masses as a function of virial mass of their hosting halos, but this time lining up the three corresponding galaxies with each other so that the comparison in stellar mass can be done on an individual rather than just on a statistical basis. The very small differences in the virial masses between corresponding galaxies are suppressed in the figure for visual clarity by adopting, for definiteness, the viral mass value of the SH model. Fig.~\ref{Fig:stellarmasses} confirms that the differences in the stellar masses of galaxies in the same halo are individually very small, with no clear systematic differences between the different models for star formation. This demonstrates once more that the cumulative stellar mass of galaxies in cosmological simulations is primarily determined by the net balance of baryonic inflows and outflows and its regulation by galaxy scale feedback processes, but not sensitively by the star formation efficiency in the cold ISM mass itself. Based only on studies of isolated galaxies, this result may be viewed as counter-intuitive, but for cosmological simulations it is commonly expected, as pointed out in previous analytic \citep[e.g.][]{FaucherGiguere2013, Carr2023} and numerical work \citep[e.g.][]{Hopkins2011, Semenov2018}. Note however that residual dependencies on the local star formation efficiency can still occur, for example if this modifies the clustering of supernovae, and then in turn the strength of galactic outflows.

\begin{figure}
  \centering
\resizebox{8.5cm}{!}{\includegraphics{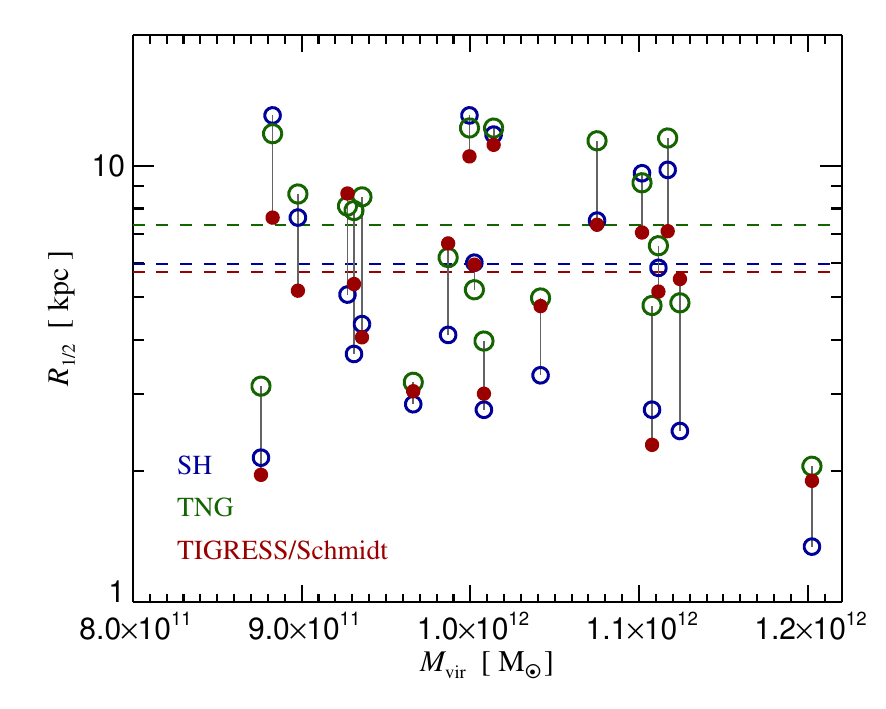}}
\caption{Stellar half-mass radii comparison for our
sample of galaxies in $10^{12}\,{\rm M}_\odot$ halos, for the SH, TNG and TIGRESS/Schmidt models, as labelled. Corresponding galaxies that are centrals in the same dark matter halos are lined up by thin vertical lines, whereas the very small differences in the virial mass are suppressed in this figure for visual clarity. Horizontal dashed lines show the mean radii  the corresponding samples of galaxies. 
  \label{Fig:stellarhalfmassradii}}
\end{figure}

\begin{figure*}
  \centering
  \resizebox{8.5cm}{!}{\includegraphics{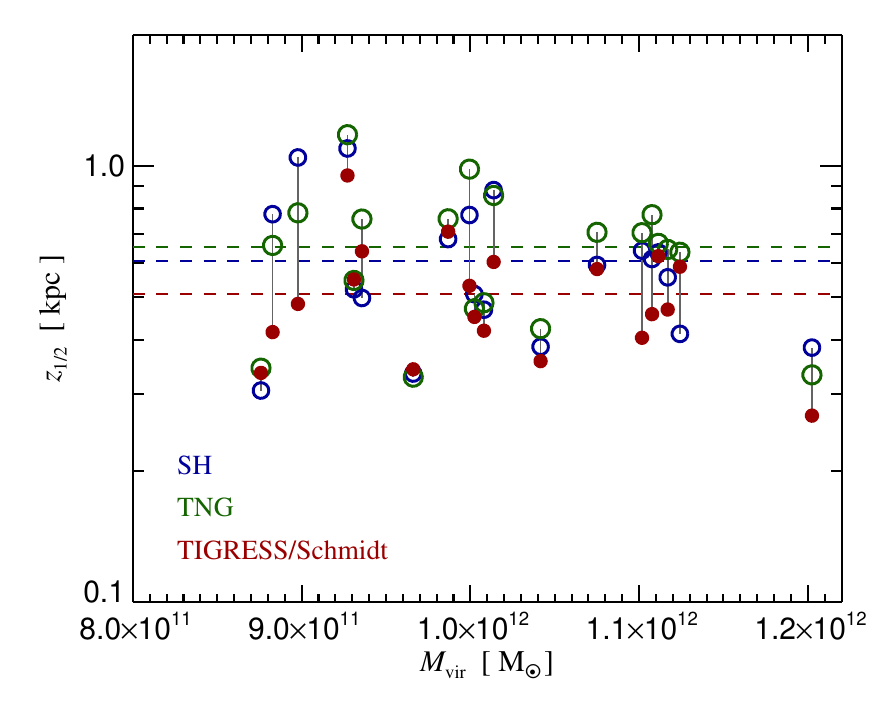}}%
  \resizebox{8.5cm}{!}{\includegraphics{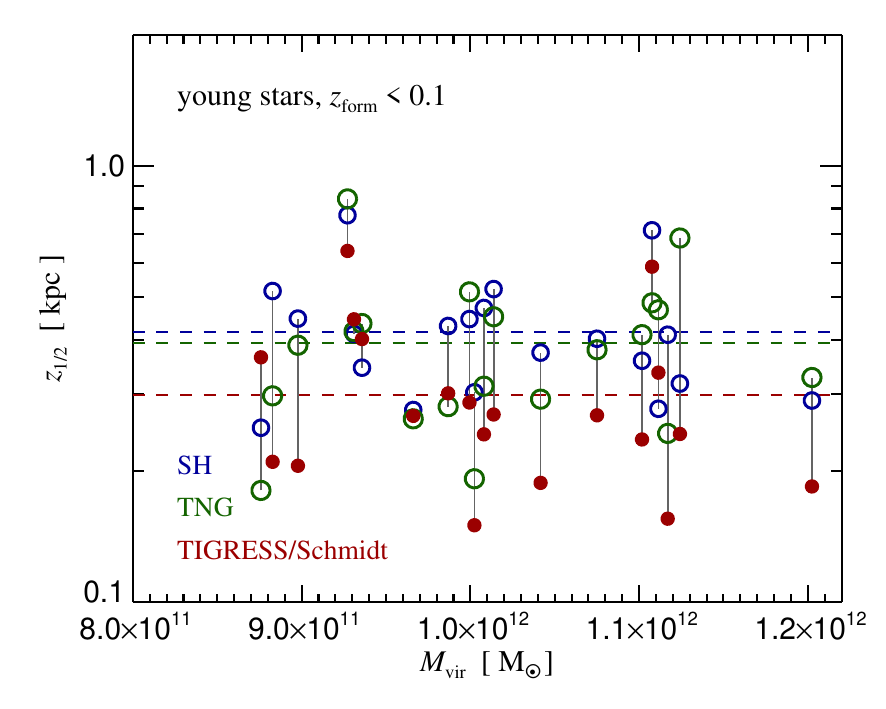}}%
\caption{Detailed object-to-object comparison of the stellar half-mass scale-heights in our cosmological galaxy sample at virial halo mass scale $10^{12}\,{\rm M}_\odot$. All galaxies are analysed at redshift $z=0$ and have been separately simulated with the SH, TNG, and TIGRESS/Schmidt models. Averages over the samples of 20 disk galaxies are shown as dashed horizontal lines. As the left panel shows, the scale heights of TIGRESS/Schmidt are  smaller than SH and TNG, even so the measured values are considerably larger than the intrinsic thickness of the star-forming layers at birth. The right panel shows the same measurement but restricted to young stars formed at redshifts $z<1$. In this case, much smaller values are inferred, showing that the full set of star particles is affected by substantial heating processes during cosmic evolution, and may also contain a substantial spheroidal contribution in some of the objects.
  \label{Fig:stellarscaleheight}}
\end{figure*}

\subsection{Differences in galaxy radii and scale heights}

We now turn to an investigation of how the main structural properties of these galaxies compare. Again, we focus on the $10^{12}\,{\rm M}_\odot$ sample for simplicity. In Figure~\ref{Fig:stellarhalfmassradii}, we look at the stellar half-mass radii, in a matched comparison as carried out before for the stellar masses. Here we find a signal that on average the TIGRESS/Schmidt galaxies are the most concentrated, which is consistent with the slightly stepper expected relation between the surface densities of star formation and gas density which we saw in Fig.~\ref{Fig:sfrvsgas}. However, the effect is small and not particularly pronounced, as this difference only clearly shows up with respect to TNG and not to the SH model. This suggests that differences in galaxy sizes due to the different model variations are minor overall, and are not significant given other sources of uncertainties and scatter.

Next we look at the scale-heights of the stellar disks, measured in the
form of a half-mass thickness as before, i.e.~50 percent of the stellar mass is contained within $|z| < z_{1/2}$. No attempt has been made to separate out a possibly present bulge component. In Figure~\ref{Fig:stellarscaleheight} we show the corresponding results, again for the $10^{12}\,{\rm M}_\odot$ set aligned in a matched fashion. The left panel shows results for all the stars at $z=0$. We find a clear systematic difference in the average thickness of the galaxies, with the TIGRESS/Schmidt galaxies being about 20\% thinner than TNG and SH. This is consistent with the general expectation from isolated  disk galaxy simulations, but also quantitatively smaller than may have been expected based on our isolated galaxies results. Note also that the absolute values are quite a bit larger than the native thickness we expect for stars formed in star-forming layers of disk galaxies. This indicates the presence of substantial heating effects that puff up the thickness of the stellar disks, including processes such as spiral arms and bars, interactions and mergers with infalling satellites, radial migration, and potential fluctuations due to bursty wind feedback \citep{Grand2016}. There could also be numerical heating effects related to two-body interactions with massive dark matter particles \citep{Ludlow2023}, or scattering on dense gaseous lumps.

An indication that such secular heating processes are at play is given by the right panel of Figure~\ref{Fig:stellarscaleheight}, in which we restrict the measurement of the thickness to young stars that have formed at redshifts $z<0.1$. In this case, we measure values that are nearly a factor 2 smaller for all the models, and the relative difference between TIGRESS/Schmidt and TNG/SH has slightly increased as well. But also here, the intrinsic thickness is mostly forgotten due to the strong influence of disk heating effects. In addition, of course, our measurement is relatively coarse and does not exclude stars that are not in the disk in the first place. Also, it neglects effects such as warps and other disk corrugations \citep{Gomez2017} that are known to  be frequent and often quite strong in cosmological simulations. Finally, note that -- perhaps somewhat counter-intuitively -- an excessively thin height of a stellar disk when it is born may in fact create thicker stellar distributions in the long run, simply because such disks are prone to develop instabilities that lead to buckling or fragmentation of the disk into lumps. We leave a further investigation of the issue of disk thickness, and whether stellar particles formed in the TIGRESS/Schmidt model should be endowed with a random velocity component upon birth, to a future investigation.

\section{Summary and conclusions} \label{SecConc}

We have implemented a new subgrid equation of state for the regulation of star formation in the ISM of cosmological simulations. They have too low resolution to spatially resolve the real multi-phase structure of the ISM, which in the past led to the concept of explicit subgrid models that try to model this critical component of galaxies as a single-phase that is governed by an effective average pressure. The relation between the mean average density and the average pressure is referred to as equation of state. SH had introduced such a model two decades ago based on a coarse analytic description of the multiphase structure of the ISM. Subsequently, many versions of similar models,  often based on heuristic arguments, have been used in cosmological simulations, and the associated reduction in computational cost has been instrumental for being able to carry out large-volume hydrodynamic cosmological simulations of galaxy formation in the first place.

In recent times, simulations that directly resolve star formation in the ISM and take most of the relevant physics into account have seen tremendous progress. Importantly, they now make it possible to place the concept of an effective equation of state onto a sound quantitative basis by allowing a direct measurement of the relation between mean density and mean pressure, and furthermore, the resulting star formation rates for a given surface density in disk galaxies. We here have taken the results from the TIGRESS-classic simulation suite to construct a new variant of a subgrid model that is directly calibrated against the results of these high-resolution simulations, and we have compared it to the older TNG and SH models. 

In particular, we have replaced the equation of state previously used in the TNG simulations with the TIGRESS-classic result. We have also implemented a new star formation law that we call TIGRESS/Schmidt, which to good accuracy reproduces the relation measured in TIRGRESS for the mid-plane pressure and the surface density of star formation in disk galaxies. The outcome is not identical to the star formation expected based on the pressure-regulated, feedback-modulated (PRFM) theory when a very massive and thin stellar disk is present, but it is very close when the gas dominates the vertical gravity. An important practical advantage of TIGRESS/Schmidt is that it is considerably easier to implement in a cosmological simulation as it is formulated based only on the local gas properties of individual Voronoi cells. We have therefore focused  here on studying first results for the TIGRESS/Schmidt scenario but will analyse the differences between TIGRESS/Schmidt and PRFM more closely in a companion paper.

For our analysis of differences between TIGRESS/Schmidt, TNG and SH in galaxy simulations we first considered a small sample of isolated disk galaxies which we mostly used to verify our theoretical expectations for vertical structure, scaling relations, and stability in the different models. We have then turned to cosmological zoom-in simulations of galaxies re-simulated from a very large 740~Mpc box taken from the MillenniumTNG simulation suite. To this end we have introduced a novel initial conditions code that supports what we call `multi zoom-in' simulations in which an arbitrary set of target halos (not just one as in the traditional version of this approach) can be specified for resimulation at higher resolution.  We have chosen three different narrow mass bins, centred around a virial mass of $10^{12}\,{\rm M}_\odot$, and factors of 8 higher and lower. For each of these mass ranges, we randomly selected 20 halos from the parent simulation and computed them with the SH, TNG and TIGRESS/Schmidt again at a number of different numerical resolutions, with the goal to robustly identify the most important differences resulting from the different treatment of the equation of state and the star-formation law. Importantly, additional feedback prescriptions in the form of galactic winds and supermassive black holes, taken from the IllustrisTNG treatment, were applied identically in all three simulation models. Our main findings can be summarized as follows:

\begin{itemize}
\item The equation-of-state pressure predicted by the TIGRESS-classic simulations is generally lower than that used in TNG, let alone than in the very stiff original SH model. The EOS slope is also a bit softer in TIGRESS-classic, at least in the low density regime, but still sufficient to stabilize the gas against numerical fragmentation when a Lagrangian code and a sufficiently high gas mass resolution is used.

\item The difference in the equation-of-state implies that TIGRESS-classic models have substantially thinner gas disks  in star-forming disk galaxies compared to TNG and SH, and this naturally creates thinner stellar disks as well. At equal surface densities, this also lowers the Toomre-$Q$ stability parameter of gas disks, implying that high surface density disks in the TIGRESS/PRFM model are more prone to axisymmetric instabilities and local clump formation. 

\item The relation between gas surface density and star formation rate surface density  for isolated disk galaxies (the Kennicutt relation) is slightly steeper in TIGRESS/Schmidt than in TNG and SH, but remains fairly close in amplitude overall. This can create slightly smaller disk scale lengths in TIGRESS/Schmidt.

\item When analysing star formation in isolated galaxies, we find that our default strategy for initializing the velocities of newly created star particles -- which is simply that star particles inherit the velocity of the parent gas cell -- leads to stellar disks that are substantially thinner, by up to a factor of two, than the vertical distribution of star formation itself. This is because all stars start with negligible  vertical velocity dispersion in this case, and they are thus likely to be found always at smaller $z$-coordinate at any given later time. It may be preferable to impose a suitable thermal velocity dispersion onto newly created star particles, such that they can attain a Jeans equilibrium with a larger thickness already at birth. 

\item Despite the differences in the equation of state and the star-formation prescriptions, cosmological simulations of galaxies with the TIGRESS/Schmidt, TNG, and SH models show a close convergence of their stellar masses. This can be understood as a result of the strong self-regulation of global star formation by the wind feedback model that has been included in these simulations; the stellar mass is hardly influenced by the local star formation efficiency because the amount of baryons that makes it into the cold reservoir available for star formation is regulated by halo-scale cooling and heating by galactic wind feedback. The latter part of the modelling is identical in the TIGRESS/Schmidt, TNG, and SH simulations studies in this work. An important goal of current work in the Learning the Universe collaboration is to replace this wind prescription with the numerical model ARKENSTONE \citep{Smith2024a, Smith2024b, Benett2024} for multiphase outflows, and calibrate it  directly against the winds measured in TIGRESS-like simulations \citep{Kim2020winds}.

\item The TIGRESS/Schmidt model tends to produce much clumpier disk morphologies than SH and TNG. This reflects the higher susceptibility of the thin gas disks in TIGRESS/Schmidt to various types of instabilities. This is reinforced  by the even thinner stellar disks that are formed in this model due to the dynamical coldness of the formed stellar distribution at birth. It will be an interesting question to test whether the frequency of clumps in TIGRESS/Schmidt is significantly reduced once this aspect of the modelling is modified. We note that while some clumpy disks are observed, there are also many smooth disks seen out to high redshift, and it could well be that TIGRESS/Schmidt yields a good match once this is addressed.

\end{itemize}

In forthcoming work, we will carry out a more quantitative analysis of the morphological trends seen in the galaxies of our zoom-in simulations. We will also investigate potential differences between TIGRESS/Schmidt and the application of a star formation law that is fully and not only approximately consistent with the PRFM theory. Finally, we will analyse results from uniformly resolved cosmological box simulations with the PRFM model, and carry out an analysis similar to that made by \citet{Hassan2024}  for the TNG50 simulation in postprocessing. This will check whether our theoretical understanding of the regulation of star formation as obtained from high-resolution ISM simulations is correctly reflected by our new effective sub-grid models when applied to cosmological simulations. If the answer turns out to be affirmative, this can be viewed as an important confirmation that the results of cosmological simulations have a solid physical grounding despite the sweeping approximations of the galaxy formation physics they need to make.


\section*{Acknowledgements}
JB and VS acknowledge support by the Simons Collaboration on ``Learning the Universe''. ECO and C-GK acknowledge support from the Simons Foundation under grant 888968. RW acknowledges funding of a Leibniz Junior Research Group (project number J131/2022). The Flatiron Institute is supported by the Simons Foundation.

\section*{Data Availability}
The data underlying this article will be shared upon reasonable request to the corresponding authors.




\bibliographystyle{mnras}
\bibliography{main}




\bsp	
\label{lastpage}
\end{document}